\documentclass[11pt,a4paper]{article}

\usepackage[utf8x]{inputenc}
\usepackage[T1]{fontenc}
\usepackage{color}
\usepackage{amssymb}
\usepackage{amsmath,amsfonts}
\usepackage{graphicx}
\usepackage{mathtools}
\usepackage{ragged2e}
\usepackage{psfrag}
\usepackage{relsize}
\usepackage{upgreek}
\usepackage{booktabs}
\usepackage{mathabx}

\usepackage{bm}
\usepackage{mathrsfs}
\usepackage{graphicx}
\usepackage{verbatim} 
\usepackage[english]{babel}
\usepackage[hypertexnames=false]{hyperref} 
\hypersetup{
citecolor=red,
colorlinks=true,
filecolor=red,
linkcolor=blue,
linktocpage=true,
urlcolor=blue
} 
\usepackage[titletoc,toc,title]{appendix}
\numberwithin{equation}{section}
\usepackage{cite}
\usepackage{epsfig}
\usepackage{float}
\usepackage{enumitem}
\usepackage[font={footnotesize,it}]{caption}
\usepackage{cases}
\usepackage{tikz,empheq}

\newcommand{\spll}{{/\kern-0.2em/}}
\newcommand\trick[1]{}
\newcommand{\be}{\begin{equation}} 
\newcommand{\ee}{\end{equation}}
\newcommand{\eq}[1]{(\ref{#1})}
\newcommand{\bit}{\begin{itemize}}  \newcommand{\eit}{\end{itemize}}
\newcommand{\ben}{\begin{enumerate}}  \newcommand{\een}{\end{enumerate}}

\newcommand{\bra}[1]{\langle #1|}
\newcommand{\ket}[1]{|#1 \rangle}
\newcommand{\mel}[3]{\langle #1\vert #2\vert #3\rangle}

\newcommand{\rf}[1]{(\ref{#1})}

\def\bd{\begin{document}}
\def\ed{\end{document}}
\def\bea{\begin{eqnarray}}
\def\eea{\end{eqnarray}}

\def\la{\langle}
\def\ra{\rangle}

\newcommand{\jj}[6]{\begin{pmatrix}#1&#2&#3\\#4&#5&#6\end{pmatrix}}
\newcommand{\sixj}[6]{\begin{Bmatrix}#1&#2&#3\\#4&#5&#6\end{Bmatrix}}
\newcommand{\ninej}[9]{\begin{Bmatrix}#1&#2&#3\\#4&#5&#6\\#7&#8&#9\end{Bmatrix}}
\newcommand{\CG}[6]{\left\langle #1\,#2;#3\,#4\middle|#5\,#6\right\rangle}

\newcommand{\Khat}{\widehat{\mathcal K}}
\newcommand{\Vhat}{\widehat{\mathcal V}}
\newcommand{\What}{\widehat{\mathcal W}}
\newcommand{\inner}[2]{\left\langle #1,#2\right\rangle}
\newcommand{\Order}{\mathcal O}

\def\sst{\scriptscriptstyle}
\def\thetabar{\bar\theta}
\def\Tr{{\rm Tr}}
\def\dd{\mathrm d}
\def\XN{X^{(N)}}
\def\XNp{X^{(N')}}
\def\An{A^{(n)}} 
\def \lp{l_P}

\newcommand{\Leff}{\Lambda^{\rm eff}}
\newcommand{\Lcl}{\Lambda^{\rm cl}}
\newcommand{\MP}{M_{\!P}}
\newcommand{\Mzero}{M_0}
\newcommand{\HH}{\mathrm H}
\newcommand{\EE}{\mathrm E}
\newcommand{\one}{\mathbf 1}
\newcommand{\vect}[1]{\bm{#1}}

\def\a{\alpha}      \def\da{{\dot\alpha}}  \def\dA{{\dot A}}
\def\b{\beta}       \def\db{{\dot\beta}}
\def\g{\gamma}  \def\G{\Gamma}  \def\dc{{\dot\gamma}}
\def\d{\delta}  \def\D{\Delta}  \def\ddt{\dot\delta}
\def\e{\epsilon}
\def\ve{\varepsilon}
\def\uve{\upvarepsilon}
\def\f{\phi}    \def\F{\Phi}    \def\vvf{\f}
\def\vphi{\varphi}
\def\h{\eta}
\def\k{\kappa}
\def\l{\lambda} \def\L{\Lambda}
\def\m{\mu} \def\n{\nu}
\def\o{\omega}\def\O{\Omega}
\def\p{\pi} \def\P{\Pi}
\def\r{\rho}
\def\s{\sigma}  \def\S{\Sigma}
\def\t{\tau}
\def\th{\theta} \def\Th{\Theta} \def\vth{\vartheta}
\def\X{\Xeta}
\def\z{\zeta}

\def\na{\nabla}

\def\cA{{\cal A}} \def\cB{{\cal B}} \def\cC{{\cal C}}
\def\cD{{\cal D}} \def\cE{{\cal E}} \def\cF{{\cal F}}
\def\cG{{\cal G}} \def\cH{{\cal H}} \def\cI{{\cal I}}
\def\cJ{{\mathscr J}} \def\cK{{\cal K}} \def\cL{{\cal L}}
\def\cM{{\cal M}} \def\cN{{\cal N}} \def\cO{{\cal O}}
\def\cP{{\cal P}} \def\cQ{{\cal Q}} \def\cR{{\cal R}}
\def\cS{{\cal S}} \def\cT{{\cal T}} \def\cU{{\cal U}}
\def\cV{{\cal V}} \def\cW{{\cal W}} \def\cX{{\cal X}}
\def\cY{{\cal Y}} \def\cZ{{\cal Z}}
\def\ccb{{\cal b}}
\def\ct{{\cal t}}

\def\ua{\underline{\alpha}}
\def\uc{\underline{\phantom{\alpha}}\!\!\!\gamma}
\def\um{\underline{\mu}}
\def\ud{\underline\delta}
\def\ue{\underline\epsilon}
\def\una{\underline a}\def\unA{\underline A}
\def\unb{\underline b}\def\unB{\underline B}
\def\unc{\underline c}\def\unC{\underline C}
\def\und{\underline d}\def\unD{\underline D}
\def\une{\underline e}\def\unE{\underline E}
\def\unf{\underline{\phantom{e}}\!\!\!\! f}\def\unF{\underline F}
\def\unm{\underline m}\def\unM{{\underline M}}
\def\unn{\underline n}\def\unN{{\underline N}}
\def\unp{\underline{\phantom{a}}\!\!\! p}\def\unP{\underline P}
\def\unq{\underline{\phantom{a}}\!\!\! q}
\def\unQ{\underline{\phantom{A}}\!\!\!\! Q}
\def\unH{\underline{H}}

\def\As {{A \hspace{-6.4pt} \slash}\;}
\def\bs {{b \hspace{-6.4pt} \slash}\;}
\def\Ds {{D \hspace{-6.4pt} \slash}\;}
\def\Gts {{\Gt \hspace{-6.4pt} \slash}\;}
\def\ds {{\del \hspace{-6.4pt} \slash}\;}
\def\ss {{\s \hspace{-6.4pt} \slash}\;}
\def\ks {{ k \hspace{-6.4pt} \slash}\;}
\def\ps {{p \hspace{-6.4pt} \slash}\;}
\def\xs {{x \hspace{-6.4pt} \slash}\;}
\def\pas {{{p_1} \hspace{-6.4pt} \slash}\;}
\def\pbs {{{p_2} \hspace{-6.4pt} \slash}\;}
\def\cFs {{{\cal F} \hspace{-6.4pt} \slash}\;}
\def\Dss {{D \hspace{-7.5pt} \slash}\;}
\def\dss {{\del \hspace{-7.0pt} \slash}\;}

\def\Ah{{\hat{A}}}
\def\Bh{{\hat{B}}}
\def\Dh{{\hat{D}}}
\def\Gh{{\hat{G}}}
\def\Fh{{\hat{F}}}
\def\Ih{{\hat{I}}}
\def\Jh{{\hat{J}}}
\def\Kh{{\hat{K}}}
\def\Lh{{\hat{L}}}
\def\Ph{{\hat{P}}}
\def\Rh{{\hat{R}}}
\def\Vh{{\hat{V}}}
\def\Xh{{\hat{X}}}
\def\Yh{{\hat{Y}}}

\def\hn{{\hat{n}}}

\def\ah{{\hat{\a}}}
\def\bh{{\hat{\b}}}
\def\gh{{\hat{\g}}}
\def\dh{{\hat{\d}}}
\def\lh{{\hat{\l}}}
\def\rh{{\hat{\r}}}
\def\hh{\hat{h}}
\def\uh{\hat{u}}
\def\xh{\hat{x}}
\def\yh{\hat{y}}
\def\ph{\hat{p}}
\def\xih{\hat{\xi}}
\def\chih{\hat{\chi}}
\def\Psih{\hat{\Psi}}
\def\phih{\hat{\phi}}

\def\psit{\tilde{\psi}}
\def\Psit{\tilde{\Psi}}
\def\Psibt{\tilde{\bar{Psi}}}

\def\lambdat{\tilde {\lambda}}
\def\st{\tilde{\sigma}}

\def\delt{\tilde{\delta}}
\def\Phit{\tilde{\Phi}}
\def\Phitb{\overline{\tilde{Phi}}}
\def\tht{\tilde{\th}}
\def\lt{\tilde{\l}}
\def\chit{\tilde{\chi}}
\def\phit{\tilde{\phi}}

\def\At{\tilde{A}}
\def\Bt{\tilde{B}}
\def\Ct{\tilde{C}}
\def\Dt{\tilde{D}}
\def\Et{\tilde{E}}
\def\Ft{\tilde{F}}
\def\Gt{\tilde{G}}
\def\Ht{\tilde{H}}
\def\It{\tilde{I}}
\def\Jt{\tilde{J}}
\def\Pt{\tilde{P}}
\def\Ot{\tilde{O}}
\def\Mt{\tilde{M }}
\def\Nt{\tilde{N}}
\def\St{\tilde{S}}
\def\Vt{\tilde{V}}
\def\Xt{\tilde{X}}
\def\at{\tilde{a}}
\def\ct{\tilde{c}}
\def\dt{\tilde{d}}
\def\htt{\tilde{h}}
\def\ft{\tilde{f}}
\def\gt{\tilde{\gamma}}
\def\pt{\tilde{p}}
\def\qt{\tilde{q}}
\def\rt{\tilde{r}}
\def\tt{{\tilde{t}}}
\def\nt{\tilde{n}}
\def\ut{\tilde{u}}
\def\wt{\tilde{w}}
\def\zt{\tilde{z}}
\def\xt{\tilde{x}}
\def\yt{\tilde{y}}
\def\Psit{\tilde{\Psi}}
\def\phit{\tilde{\phi}}
\def\tD{\tilde{\D}}

\def\eb{\bar{\epsilon}}
\def\delb{\bar{\partial}}
\def\thb{\bar{\theta}}
\def\mub{\bar{\mu}}
\def\lamb{\bar{\l}}
\def\psib{\bar{\psi}}
\def\sb{\bar{\sigma}}
\def\xib{\bar{\xi}}
\def\chib{\bar{\chi}}

\def\Psib{\bar{\Psi}}
\def\Phib{\bar{\Phi}}
\def\Lamb{\bar{\Lambda}}
\def\Sb{{\overline \Sigma}}
\def\hb{\bar{h}}
\def\qb{\bar{q}}
\def\wb{\bar{w}}
\def\ub{\bar{u}}
\def\zb{{\bar{z}}}
\def\Hb{\bar{H}}
\def\Qb{{\bar Q}}
\def\ob{\overline{\omega}}

\def\Ab{{\overline A}} \def\Bb{{\overline B}}
\def\Db{{\overline D}} \def\Eb{{\overline E}} \def\Fb{{\overline F}}
\def\Gb{{\overline G}}
\def\Ib{{\overline I}}
\def\Jb{{\overline J}} \def\Kb{{\overline K}} \def\Lb{{\overline L}}
\def\Mb{{\overline M}} \def\Nb{{\overline N}} \def\Ob{{\overline O}}
\def\Pb{{\overline P}}  \def\Rb{{\overline R}}
 \def\Tb{{\overline T}} \def\Ub{{\overline U}}
\def\Vb{{\overline V}} \def\Wb{{\overline W}} \def\Xb{{\overline X}}
\def\Yb{{\overline Y}} \def\Zb{{\overline Z}}

\def\fb{{\overline f}}
\def\gb{{\overline g}}
\def\nb{{\overline n}}
\def\mb{{\overline m}}
\def\lb{{\overline l}}
\def\yb{{\overline y}}

\def\ldel{{\overleftarrow{\del}}}
\def\rdel{{\overrightarrow{\del}}}
\def\ldeldel{{\overleftarrow{\del^2}}}
\def\rdeldel{{\overrightarrow{\del^2}}}
\def\ldelb{{\overleftarrow{\bar{\del}}}}
\def\rdelb{{\overrightarrow{\bar{\del}}}}

\def\ba{{\bf a}}
\def\bk{{\bf k}}
\def\bl{{\bf l}}
\def\bp{{\bf p}}
\def\bq{{\bf q}}

\def\br{{\bf r}}
\def\bt{{\bf t}}
\def\bu{{\bf u}}
\def\bv{{\bf v}}
\def\bx{{\bf x}}
\def\by{{\bf y}}
\def\bA{{\bf A}}
\def\bR{{\bf R}}
\def\bV{{\bf V}}

\def\bz{{\boldsymbol{\zeta}}}

\def\bone{{\bf 1}}

\def\va{{\vec a}}
\def\vj{{\vec j}}
\def\vk{{\vec k}}
\def\vp{{\vec p}}
\def\vq{{\vec q}}
\def\vx{{\vec x}}
\def\vy{{\vec y}}
\def\vu{{\vec u}}
\def\vv{{\vec v}}
\def \vB{{\vec B}}
\def \vE{{\vec E}}
\def \vH{{\vec H}}
\def \vg{{\vec g}}

\def\vs{{\vec \sigma}}
\def\vtau{{\vec \tau}}

\newcommand{\ov}[1]{\overrightarrow{#1}}

\def\frA{\mathfrak{A}}
\def\frB{\mathfrak{B}}
\def\frC{\mathfrak{C}}
\def\frD{\mathfrak{D}}
\def\frE{\mathfrak{E}}
\def\frF{\mathfrak{F}}
\def\frG{\mathfrak{G}}
\def\frH{\mathfrak{H}}
\def\frM{\mathfrak{M}}
\def\frN{\mathfrak{N}}
\def\frR{\mathfrak{R}}
\def\frW{\mathfrak{W}}

\def\fra{\mathfrak{a}}
\def\frb{\mathfrak{b}}
\def\frf{\mathfrak{f}}
\def\frg{\mathfrak{g}}
\def\frh{\mathfrak{h}}
\def\frl{\mathfrak{l}}
\def\frs{\mathfrak{s}}
\def\fri{\mathfrak{i}}
\def\frj{\mathfrak{j}}

\def\ma{\mathfrak{a}}
\def\mb{\mathfrak{b}}
\def\mg{\mathfrak{g}}
\def\mh{\mathfrak{h}}
\def\mB{\mathfrak{B}}
\def\mR{\mathfrak{R}}
\def\mN{\mathfrak{N}}

\def\Xk{X^{(k)}}
\def\Xkp{X^{(k')}}
\def\psik{\psi^{(k)}}
\def\psikp{\psi^{(k')}}
\def\Nk{N_k}
\def\Nkp{N_{k'}}

\newcommand{\nn}{{\nonumber}}

\def\d{\delta}\def\D{\Delta}

\def\pa{\partial} \def\del{\partial}
\def\xx{\times}
\def\uno{\mbox{1 \kern-.59em {\rm l}}}

\def\trp{^{\top}}
\def\inv{^{-1}}
\def\dag{\dagger}
\def\pr{^{\prime}}

\def\rar{\rightarrow}
\def\lar{\leftarrow}
\def\lrar{\leftrightarrow}

\newcommand{\0}{\,\!}      
\def\im{\imath}
\def\jm{\jmath}

\newcommand{\tr}{\mbox{tr}}
\newcommand{\slsh}[1]{/ \!\!\!\! #1}

\newcommand{\1}{\mbox{1}\hspace{-0.25em}\mbox{l}}

\def\vac{|0\rangle}
\def\lvac{\langle 0|}

\def\hlf{\frac{1}{2}}
\def\ove#1{\frac{1}{#1}}
\newcommand{\hot}[1]{\frac{#1}{2}}

\def\Box{\square}
\def\CC {\mathbb{C}}
\def\FF {\mathbb{F}}
\def\RR{\mathbb{R}}
\def\NN{\mathbb{N}}
\def\ZZ{\mathbb{Z}}
\def\bb#1{{\bf #1}}
\def\bcomment#1{}
\def\bfhat#1{{\bf \hat{#1}}}
\def\VEV#1{\left\langle #1\right\rangle}

\newcommand{\ex}[1]{{\rm e}^{#1}} \def\ii{{\rm i}}

\newcommand{\lrbrk}[1]{\left(#1\right)}
\newcommand{\lrsbrk}[1]{\left[#1\right]}
\newcommand{\sfrac}[2]{{\textstyle\frac{#1}{#2}}}

\def\stw{{\sqrt{2}}}

\def\rf {{\rm f}}
\def\ri {{\rm i}}
\def\rj {{\rm j}}
\def\rn {{\rm n}}
\def\rk {{\rm k}}
\def\rl {{\rm l}}
\def\rr {{\rm r}}

\def\rQ {{\scriptscriptstyle \rm \cQ}}
\def\rR {{\scriptscriptstyle \rm \cR}}

\def\cQb{{\cal \Qb}}
\def\cRb{{\cal \Rb}}
\def\cWb{{\cal \Wb}}

\def\fd {{\rm N}}
\def\afd {{\overline{\rm N}}}

\def \II {I\hspace{-.1em}I\hspace{.1em}}
\def \IIA {\mbox{\II A\hspace{.2em}}}
\def \IIB {\mbox{\II B\hspace{.2em}}}
\def \gs {g^s}
\def \ls {\lambda^s}

\def \I {{\cal I}}
\def \qs {q\hspace{-.53em}/\hspace{.15em}}
\def \ks {k\hspace{-.53em}/\hspace{.15em}}
\def \YM {{\mbox{\tiny YM}}}
\def \gym {g_{\YM}}

\def \Lc {\L_c}
\def\IR{\relax{\rm I\kern-.18em R}}
\def \id {{\bf 1}}

\def\cci{\ell}
\def\ccj{\ell'}

\def\bbq{\pmb{q}}
\def\bom{\pmb{\o}}
\def\bJ{\pmb{J}}
\def\bM{\pmb{M}}
\def\bB{\pmb{B}}
\def\bn{\pmb{n}}
\def\bE{\pmb{E}}

\newcommand{\rrr}[1]{\vskip 0.2cm \noindent{\bf #1} ---}

\newcommand{\KK}{\mathcal K}

\long\def\symbolfootnote[#1]#2{\begingroup%
\def\thefootnote{\fnsymbol{footnote}}\footnote[#1]{#2}\endgroup}
\long\def\RemarkBox#1{\begin{flushleft}\fbox{\begin{minipage}
{17.5cm}{\bf Remark:} ~#1\end{minipage}}\end{flushleft}}

\newcommand{\nthu}{{\it Department of Physics, National Tsing-Hua University,
    Hsinchu 30013, Taiwan}}

\newcommand{\ctc}{{\it
Center for Theory-Computation-Data Science Research, 
National Tsing-Hua University, Hsinchu 30013, Taiwan}}

\newcommand{\ncts}{{\it Physics Division,
    National Center for Theoretical Sciences, Taipei 10617, Taiwan}}

\begin{document}
\begin{center}
\vspace{20pt}
  
\thispagestyle{empty}
              {\Large \bf Quantum Stability of the  Fuzzy Sphere Black Hole
                Horizon}
               
\vspace{25pt}

Chong-Sun Chu

\vspace{0.2cm}              

\vspace{5pt}\nthu\\
\vspace{5pt}\ctc\\
\vspace{5pt}\ncts

\vspace{1cm}

\begin{abstract}

We establish the perturbative stability of the fuzzy-sphere black-hole
horizon in the large-$N$ matrix quantum mechanics of
\cite{Chu:2024qil}. Large-$N$ counting shows that the leading quantum
correction to the fluctuation spectrum is the planar bosonic one-loop
contribution, while higher-loop bosonic effects and the leading
fermionic one-loop contribution are parametrically suppressed.  The
classical spectrum contains a tachyonic $(1,2)$ mode and a marginal
$(2,3)$ mode; we show that both acquire positive quantum curvature and
are stabilized. For the general spectrum, we find that a factorization
structure of the Hessian controls the leading quantum curvature and
renders it non-negative. Quantum effects dominate the stabilization of
generic low-angular-momentum modes, whereas classical curvature
dominates at high angular momentum.  The two effects become comparable
in the crossover regime $L \sim \sqrt{N}$, where both must be
retained.  Under smoothness and non-saturation assumptions supported
by finite-$N$ numerical tests, positivity of the quantum fluctuation
spectrum is thus established for sufficiently large but finite
$N$. This local curvature stability is compatible with
non-perturbative monopole tunneling. Intriguingly, the same
$L\sim\sqrt{N}$ mesoscopic angular momentum range also dominates the
tunneling process, suggesting a distinguished IR--UV crossover regime
in the quantum dynamics of the black-hole horizon.

  
\end{abstract}
\end{center}

\newpage
\setcounter{footnote}{0}

\tableofcontents

\section{Introduction}

Recently a microscopic model of a quantum black hole has been proposed
\cite{Chu:2024qil}.  Inspired by the BFSS matrix model
\cite{Banks:1996vh}, the model is given by a nonsupersymmetric quantum
mechanics of large $N$ matrices, which are interpreted as
noncommutative quantized coordinates of quantum gravity.  In contrast
to BFSS, the absence of supersymmetry allows for a negative bosonic
mass term and fundamental fermions instead of adjoint fermions.  The
model admits a fuzzy sphere as a nontrivial classical solution, whose
energy and size reproduce the Schwarzschild mass--radius relation.
Quantization of the fermions on the fuzzy sphere gives rise to a
number $2N^2$ of fermionic oscillators, which we call partons.  As a
fermionic oscillator, each parton can be in one of the two states,
unoccupied or occupied. Therefore, a fuzzy sphere with half of the
partonic levels occupied has a microstate degeneracy which reproduces
precisely the Bekenstein-Hawking entropy in the leading order of large
$N$ \cite{Chu:2024qil}.  The same analysis extends to the much more
complicated Kerr black hole \cite{Chu:2024edh}.  It was therefore
proposed that the fuzzy sphere with the parton states provides a
microscopic model of a quantum black hole horizon
\cite{Chu:2024qil,Chu:2024edh}.

At first sight, the negative bosonic mass term and the use of
fundamental fermions may appear somewhat unusual. The former implies
that the Abelian configuration has no ground state, while the latter
couple to the adjoint bosonic matrices through a noncommutator-type
interaction and therefore do not exhibit the usual commutator-induced
kinetic propagation characteristic of BFSS.   Because the
fermions transform in the fundamental representation of the matrix
gauge group, they give rise to partonic states with no ordinary
spatial propagation on the fuzzy sphere.  This has a natural geometric
explanation in terms of an intrinsic
Berry monopole  associated with a  nontrivial
line bundle over the sphere. As a result, the partons are
precisely the lowest-Landau-level (LLL) states of the monopole
$B$-field, whose kinetic motion is quenched \cite{Chu:2026vzj}.  If an
external electric field is applied, however, their guiding centers can
drift, producing a Hall current and ultimately furnishing the
microscopic starting point for a feature-rich quantum membrane
paradigm \cite{Chu:2026vzj,Chu:2026qom}. Remarkably, the same
fundamental fermions also play an essential role in the tunneling of
one fuzzy sphere into a smaller one \cite{Chu:2026wyh}.  Such a
transition sheds part of the monopole charge together with the
corresponding occupied Fermi states, and a nonvanishing tunneling
amplitude requires the excess fermionic states to be precisely
accounted for by the fermion zero modes of the shed monopole
configuration. For the $N$ flavors of fundamental fermions present in
the model, this zero-mode counting works out exactly, thereby
selecting the monopole tunneling path and providing precisely the
number of zero modes needed for the transition.
It is striking that these structures, which initially appear
unrelated, turn out to be  tightly interlocked:
the same fuzzy-sphere geometry and fundamental fermion content
that generate the Berry-monopole LLL description
also provide precisely the monopole index structure required by
the tunneling process.

The stability of the fuzzy sphere is therefore central. We recall that
at the classical level, the fluctuation spectrum about the
fuzzy-sphere background has been determined explicitly
\cite{Chu:2024qil}: apart from a tachyonic (1,2) mode and a marginal
(2,3) mode, all other physical modes have positive curvature.  As a
result, classical stability is not guaranteed. While it is not
difficult to show that the bosonic one-loop correction gives positive
curvature to both exceptional modes, stabilizing the tachyonic
direction and lifting the marginal one \cite{Chu:2026wyh}, this result
also makes clear that quantum effects cannot simply be regarded as a
negligible perturbation. Once quantum corrections are sufficiently
important to reverse the sign of a classically unstable direction, one
must examine their effect on the entire fluctuation spectrum,
including modes whose classical curvature is already positive. The
establishment of the local curvature
stability of the quantum fuzzy sphere is the main
subject we will address in the present paper.

The problem has a nontrivial structure because different
angular-momentum regimes are stabilized in different ways. It turns
out that for modes of sufficiently low angular momentum $L \ll
\sqrt{N}$, the classical curvature is parametrically soft and the
quantum correction can provide the dominant stabilizing effect. At
sufficiently high angular momentum $L \gg \sqrt{N}$ , on the other
hand, the positive classical curvature becomes dominant, so that
stability follows independently of the detailed sign of the quantum
correction. Between these two limits lies an intermediate regime $L
\sim \sqrt{N}$ where the classical and quantum contributions are of
comparable magnitude and must be considered together. Thus the
stability of the fuzzy sphere does not arise from a single mechanism
acting uniformly over the spectrum, but instead results from the
complementary roles of classical and quantum effects in different
angular-momentum sectors.  We establish this picture through a
large-$N$ analysis based on a factorization property of the one-loop
Hessian operator, supplemented by numerical tests of the smoothness
and non-saturation assumptions underlying the analysis. Large-$N$
counting further shows that the leading quantum correction is bosonic,
while the leading fermion-loop contributions are suppressed by one power of $N$.

It is also important to distinguish the stability established here
from the non-perturbative decay of the fuzzy sphere discussed in
\cite{Chu:2026wyh}. The result of the present analysis,
namely the absence of negative
physical fluctuation curvatures, establishes local curvature stability
of the fuzzy sphere. This does not preclude non-perturbative decay:
the monopole tunneling process provides a
distinct decay channel through which the locally stable fuzzy sphere
can nevertheless decay.
It is also interesting to note that the scaling mode $(1,0)$
is exceptional: although its one-loop-corrected
curvature remains positive, a  
one-loop tadpole is generated. Its physical interpretation will be discussed
in Sec.~4.

The plan of the paper is as follows. In Sec.~2 we review the
fuzzy-sphere solution and its classical fluctuation spectrum, with
particular emphasis on the classically unstable mode.
In Sec. 3 we analyze the one-loop effective action and show explicitly
how the bosonic quantum correction stabilizes the (1,2) and (2,3) modes.
We then develop the large-$N$ analysis
for the full fluctuation spectrum. We point out that
the quantum curvature in the low angular momentum regime
is controlled by an interesting factorization property of the fluctuations.
By separating the spectrum into different angular-momentum regimes
in which quantum effects, classical effects, or both are
responsible for stability, we establish the quantum stability of the
fuzzy sphere
at large but finite $N$. A discussion is presented in Sec. 4.
A number of appendices provide technical details of the analysis.
In particular, numerical results are presented in an appendix to
illustrate the analytic results and to provide explicit support
for the smoothness and non-saturation
assumptions underlying the large-$N$ analysis.

\section{Classical Fuzzy Sphere}
\subsection{Classical fluctuations of the fuzzy sphere}

We start with the quantum mechanical action \cite{Chu:2024qil} 
\be \label{L}
S = \int \dd t \Tr \left[
   \frac{1}{2a_0^2 M_P} \dot{X}_a^{2}
   + \frac{M_P}{N^2} \left([X_a,X_b]^2 + 4  X_a^{2}\right)
   + i  \psi^\dag \dot{\psi}
   - a_2 \frac{M_P}{N^2}\psi^\dag \s_a X_a  \psi \right].
\ee
The rank-$N$ fuzzy sphere $S^2_N$ background is
\be
 X_a=J_a,
 \qquad [J_a,J_b]=\ii\epsilon_{abc}J_c,
 \qquad J_aJ_a=J(J+1)\bm 1_N,
 \qquad J:=\frac{N-1}{2}.
 \label{fuzzy}
\ee
Introduce the angular momentum operator $L_a$ whose action is 
$L_a f := [J_a, f]$, a basis of functions on $S^2_N$ is given by the
fuzzy spherical harmonics $Y_{\ell \m}$ which satisfy
\be
  L^2 Y_{\ell \m}=\ell(\ell+1) Y_{\ell \m},
  \qquad
  L_3 Y_{\ell \m}=\m Y_{\ell \m}.
\qquad -\ell \leq \m \leq \ell,
\ee
for
\be \label{l-range}
0\leq \ell \leq N-1.
\ee
A vector on the fuzzy sphere is acted on by the spin one rotation generator,
\be
  (S_a)_{bc}=-i \epsilon_{abc}, \qquad [S_a,S_b] = i \e_{abc} S_c.
\ee
The spherical vector basis
\be
 \bm e_{+1}=-\frac{\bm e_1+i\bm e_2}{\sqrt2},
 \qquad
 \bm e_0=\bm e_3,
 \qquad
 \bm e_{-1}=\frac{\bm e_1-i\bm e_2}{\sqrt2}.
\ee
satisfies
\be
S_3 \bm e_q = q \bm e_q, \quad q =0, \pm 1,
\ee
and the standard Condon-Shortley convention
\be
\bm e_q^* = (-1)^q \bm e_{-q}, \qquad \bm e_q^* \cdot \bm e_{q'} = \d_{qq'}.
\ee
For a Cartesian vector $\bm V$, we can define its spherical components by
\be
V_q := \bm e_q \cdot \bm V,
\ee
then scalar product takes the form
\be
\bm A \cdot \bm B = \sum_{q=-1}^1 (-1)^q A_q B_{-q}.
\ee
For a Hermitian operator, it is
\be
V_q^\dag = (-1)^q V_{-q}.
\ee
As we will see later,  the spherical basis is useful as it
allows direct coupling with the matrix spherical harmonics
$T_{\ell \m}$
and organizes the fluctuation problem into irreducible $(L,K)$ sectors.
The total generator acting on a vector-valued matrix is then
\be
   \cJ_a=L_a+ S_a
\ee
and the vector harmonics are given, for $j = \ell, \ell\pm 1$, by
\be \label{vh}
 |\ell,j,m\rangle
 =\sum_{q=-1}^{1}
 C_q^{(\ell jm)}
 Y_{\ell,m-q} e_q, \qquad
 C_q^{(\ell jm)} :=\langle\ell,m-q;1,q|j,m\rangle.
 \ee
 Here we have displayed the index $\ell$ explicitly as we will explicitly
 refer to the range \eq{l-range}
 of $\ell$ over the fuzzy sphere.
 
 At the quadratic level, a perturbation
 $X_a = J_a + Y_a$ of the fuzzy sphere 
 has the action
\be
S[Y] = \frac{1}{2 a_0^2 M_P}\int dt \inner{Y}{\left[
    \partial_t^2 -\Omega_0^2\cN_J\right] Y},
\ee
where the Hessian operator $\cN_Z$ is defined, for a general $Z_a$, by
\be \label{cNX}
(\cN_ZY)_a :=-\Big(
[Y_b,[Z_a,Z_b]] +[Z_b,[Y_a,Z_b]+[Z_a,Y_b]] +2Y_a
\Big)
\ee
and we have introduced the inner product on vector matrices 
\be
\inner{Y_1}{Y_2}=\sum_a\Tr(Y_{1a}^\dagger Y_{2a}).
\ee
In \cite{Chu:2024qil}, we found that
the Hessian $\cN_J$ of the fuzzy sphere is
diagonalized
\be
\cN_J B_a^{LKQ}=\Lambda_{LK}B_a^{LKQ},
\ee
by the  vector fuzzy harmonics
\be\label{modes}
Y_a=q (t)B_a^{L KQ},
\qquad
\sum_a\Tr\!\left[(B_a^{L KQ})^\dagger B_a^{L KQ}\right]=1,
\ee
\be\label{modes1}
1 \leq L \leq N-1, \qquad K=L-1, L, L+1, \qquad Q =-K, \cdots, K.
\ee
Note that $L=0,K=1$ is an overall $U(1)$ shift
and is not included in \eq{modes1} since it
is absent for traceless matrices. 
Here we have used the capitalized label $(L,K,Q)$ to denote the mode number
for an external vector harmonic,
and we will use 
$(\ell, j, m)$ to label the internal loop vector-harmonic modes
that will appear later in our calculation.
The classical eigenvalues are \cite{Chu:2024qil}
\be \label{cs}
\Lambda^{\rm cl}_{L K}=
\begin{cases}
L(L+3),&K=L-1,\\[1mm]
(L+1)(L-2),&K=L+1.
\end{cases}
\ee
Note that we have not displayed the branch $K=L$ which has $\Lcl_{LK} =0$
since 
this branch of perturbations consists of gauge modes, which generate
the gauge orbit $X_a\mapsto U X_aU^\dagger$. They do not enter into
our computation below and we will not talk about them for
the rest of the paper. For convenience, we refer to the
$K = L \pm 1$ branches as the $\pm$ branches.
As a result,
the quadratic Lagrangian for the fluctuations reads 
\be
  L[q] = \frac{1}{2a_0^2 M_P} \left(
  \dot{q}^{2} - \Lcl_{LK} \O_0^2 q^2\right), \qquad
  \O_0 := \frac{2a_0 M_P}{N}. 
\ee
All the curvatures \eq{cs}
are positive
except for the two exceptional modes:
\begin{align}
  (L,K)=(1,2):&\qquad \Lambda_{12}^{\rm cl}=-2, \\
  (L,K)=(2,3):&\qquad \Lambda_{23}^{\rm cl}=0.
\end{align}

In this article, we show that these curvatures are modified at the quantum level.
For $N$ large, the leading quantum correction is
dominated by planar bosonic one-loop diagrams
and these modes are stabilized. 
Before we do that we will first determine 
 the full classical potential for these two modes.
We show that the quartic potential term is positive for both cases.
As a result, classically,
the $(1,2)$ vacuum is
Higgsed while the $(2,3)$ vacuum remains stable at the origin.

\subsection{Classical potential for $(1,2)$ and $(2,3)$ modes}
\label{classical}

The complete classical bosonic
potential can be obtained by substituting the fluctuation \eq{modes} around
the fuzzy sphere into the matrix model action.
The resulting potential is quartic in the fluctuation.

\underline{The quadrupole mode $(L, K) =(1,2)$}

The perturbation mode  $(L, K) =(1,2)$ is classically tachyonic. To determine
its full classical potential, we note from \eq{Y12Q}
that the $(1,2)$-perturbation
can be represented as
\be
X_a= J_a+q A_{ab}J_b,
\qquad
A_{ab}=A_{ba},
\qquad
\tr A=0.
\ee
Using $[X_a,X_b]
=i (\d_{ac}+ q A_{ac})(\d_{bd} + qA_{bd})\epsilon_{cde}J_e$
and $su(2)$ trace identity, it is easy to obtain the
exact potential
\be
V_{\rm cl}(q)=V_0+
\frac{\MP(N^2-1)}{12N}
\left[-4q^2 \tr A^2-4 q^3 \tr A^3+
\frac{q^4}{2}\bigl(\tr A^2\bigr)^2\right], \qquad
 V_0 = -\frac{M_P (N^2-1)}{2N}.
\label{general-potential}
\ee
For $K=2$, we have five independent components of perturbation parametrized 
by the magnetic mode number $Q=-2,-1,0,1,2$.
A convenient orthogonal real basis is
\begin{align} \label{An}
A_0&=
\begin{pmatrix}
-1&0&0\\
0&-1&0\\
0&0&2
\end{pmatrix},
\qquad
A_{1c}=\sqrt3
\begin{pmatrix}
0&0&1\\
0&0&0\\
1&0&0
\end{pmatrix}, \qquad 
A_{1s}=\sqrt3
\begin{pmatrix}
0&0&0\\
0&0&1\\
0&1&0
\end{pmatrix},
\\
A_{2c}&=\sqrt3
\begin{pmatrix}
1&0&0\\
0&-1&0\\
0&0&0
\end{pmatrix},
\qquad 
A_{2s}=\sqrt3
\begin{pmatrix}
0&1&0\\
1&0&0\\
0&0&0
\end{pmatrix}.
\end{align}
They satisfy
\be
\tr(A_A A_B)=6\delta_{AB}.
\ee
The labels $1c,1s$ and $2c,2s$ denote the real and imaginary tesseral
combinations of the complex $Q=\pm1$ and $Q=\pm2$ spherical components.
Let us consider first the $Q=0$ component. One has
\be
\tr A^2=6,
\qquad
\tr A^3=6.
\ee
and so
\be
V_{Q=0}(q)=V_0+
\frac{\MP(N^2-1)}{N}
\left(-2q^2-2q^3+\frac32q^4\right).
\label{M0-potential}
\ee
Apart from the local maximum at $q=0$, there are two nonzero minima at
\be
 q_{\pm}=\frac{3\pm\sqrt{33}}{6}.
 \ee
 Similarly one can consider the pure real $|Q| = 1$ and $|Q| = 2$ directions.
 For any one of the four normalized non-axisymmetric basis directions
 $A_{1c}, A_{1s}, A_{2c}, A_{2s}$,
the invariants are all the same
\be
\tr A^2=6,
\qquad
\tr A^3=0.
\ee
Consequently,
\be
V_{|Q|=1,2}(q)=V_0+
\frac{\MP(N^2-1)}{N}
\left(-2q^2+\frac32q^4\right).
\label{nonaxis-potential}
\ee
This potential is even under $q\mapsto -q$.  Its nonzero minima occur at
$q=\pm\sqrt{\frac23}$ with
\be
V_{|Q|=1,2}
=V_0 -\frac{2}{3}\frac{\MP(N^2-1)}{N}.
\ee

\noindent\underline{The octupole mode $(L,K) =(2,3)$}

The  mode  $(L, K) =(2,3)$ has a vanishing quadratic classical potential.
This mode is however not a gauge mode and it is not exactly flat.
Here we determine its full classical potential. It is easy
to see that the cubic term vanishes due to rotational symmetry. In fact, 
consider a
general perturbation from this multiplet
\be \label{xx23}
\d X_a = \sum_{Q=-3}^3 q_Q A_a^{(3Q)}, \qquad (q_Q)^* = (-1)^Q q_{-Q}.
\ee
A cubic invariant constructed from this multiplet will have to be proportional
to the singlet contraction
\be
\sum_{Q_1,Q_2,Q_3} \jj{3}{3}{3}{Q_1}{Q_2}{Q_3} q_{Q_1}q_{Q_2}q_{Q_3}
\ee
This vanishes since this is manifestly symmetric under exchange of two $q$'s;
however
the Wigner $3j$ symbol acquires the factor
$(-1)^{3+3+3} =-1$ under the exchange of two columns.
As a result, the classical potential  takes the form
\be
V (q) = V_0 + c_4[A] q^4, \qquad c_4 >0.
\ee
Here $c_4>0$ since the quartic term is manifestly positive for
a Hermitian perturbation.
To get an idea of the explicit form of the potential,
let us consider the $Q=0$ component 
of the perturbation $\d X_a = q A_a$. From \eq{Y230}, we obtain that
\be
 A_1=-\{J_1,J_3\},\qquad
 A_2=-\{J_2,J_3\},\qquad
 A_3=3J_3^2-j(j+1)\mathbf 1_N,
\ee
Semiclassically,
\be
 \delta r(\bm n)\propto n_aA_a(\bm n)
 \propto 5n_3^3-3n_3
 \propto Y_{30}(\bm n),
\ee
so it is an octupolar deformation.
Expanding the potential around \eq{xx23}, a direct
evaluation gives
\be
V_{L=2,K=3}(q)
 =V_0+
 \frac{\MP(N^2-1)(N^2-4)(65N^2-452)}{210N}\,q^4.
 \ee
 The $(2,3)$ mode is therefore a quadratically  flat but quartically
 stabilized physical deformation.

\section{Quantum Stability of the Fuzzy Sphere}

The model of a quantum black hole
will be problematic if the (1,2)-perturbation mode is tachyonic:
the fuzzy sphere would be perturbatively unstable with
a decay time scale $\sim 1/R$ and
previous results obtained on the round fuzzy-sphere background,
including the Schwarzschild mass-radius relation,
microstate entropy, Hawking radiation, and
microscopic membrane paradigm, would not be reliable.
Fortunately this is not the case as 
we will now show that the  (1,2) mode is in fact stabilized
by bosonic one-loop effects.

Let us consider the general case and determine
the quantum corrections to the classical spectrum \eq{cs}. 
Consider the
displayed configuration
\be \label{XJB}
X_a = J_a + \sum_{Q=-K}^K q_{L KQ}(t)B_a^{L KQ}
\ee
given by the fuzzy sphere deformed by the perturbation mode $(L,K)$.
Here the vector harmonic $B_a^{LKQ}$ is unit-normalized
\be
\sum_a \Tr \left[
(B_a^{LKQ})^\dag B_a^{LKQ}
  \right] =1. 
\ee
Integrating out the bosonic and fermionic fluctuations, we obtain
the quadratic effective action
\be
\G^{(2)} = \frac{1}{2} \int \frac{d\o}{2\pi}
q_{LKQ} (-\o) \left[
  \o^2 + \O_0^2 \L_{LK}^{\rm cl} + \Pi_{LK}(\o)\right]
q_{LKQ}(\o),
\ee
where $\Pi_{LK}(\o) = \Pi^B_{LK}(\o)+\Pi^F_{ LK}(\o)$
is the self-energy and receives contribution from both
bosonic loops and fermionic loops.
The shift to the
  curvature is determined by the zero frequency part
  \be 
  \Leff_{LK} = \L_{LK}^{\rm cl} + \D  \L_{LK}, \qquad  \D  \L_{LK} :=
  \frac{\Pi_{LK}(0)}{\O_0^2}.
  \ee
  Therefore we
  need to determine $\Pi_{LK}(0)$ in order
  to decide on the quantum stability of the spectrum.

\subsection{Quantum corrections to the fuzzy sphere spectrum}

  In this subsection, we
  show that at large $N$,
  the leading quantum correction to the self-energy comes
  from planar bosonic one-loop diagrams and  is of order $\Order(1/N)$.

\subsubsection{Large-$N$ expansion of the bosonic self-energy}

We are interested in  the self-energy $\Pi^B_{LK}(0)$ obtained by
integrating out the bosonic fluctuations $Y$ in the presence of the
external deformation \eq{XJB}
\be \label{XXbarY}
 X_a=\bar X_a+Y_a,
 \qquad
 \bar X_a:=J_a+qB_a .
\ee
We remark that, here, $J_a$ is the round fuzzy-sphere background,
which satisfies
the classical equations of motion, whereas $\bar X_a$ is in
general only a displaced configuration and does not  satisfy
the equations of motion for $q\neq0$. 
This causes no difficulty for the computation of the 1PI
effective action as the effective action $\Gamma[\bar X]$,
constructed by a Legendre transformation,  is precisely
defined to be an off-shell functional.
Therefore, for an arbitrary displaced configuration \eq{XXbarY},
the self-energy can be extracted from the one-loop determinant contribution
to the 1PI effective action for the retained collective coordinate,
\be
 \Gamma^{(1)}[\bar X]
 =
 \frac12 \Tr' \log D_{\bar X},
\ee
where $D_{\bar X}$ is the quadratic fluctuation operator around the configuration
$\bar X_a$ and
$\Tr'$  will be defined below (see after Eq. \eq{D012}).

It is nevertheless useful to check explicitly that there is no
mixing linear in both the external mode $qB$ and the retained
internal modes.   Consider expansion around $J_a$ with the total displacement 
\be
 Z_a=qB_a+Y_a .
\ee
Since $J_a$ satisfies the classical equations of
motion, the  quadratic part of the classical potential contains
\be
 \frac{2M_P}{N^2}
 \left[
 q^2\langle B,\mathcal N_JB\rangle
 +2q\langle B,\mathcal N_JY\rangle
 +\langle Y,\mathcal N_JY\rangle
 \right].
\ee
Because $B=B^{LKQ}$ is an eigenmode of the round-sphere Hessian
$ \mathcal N_JB^{LKQ} = \Lambda^{\rm cl}_{LK}B^{LKQ}$,
the mixed term is proportional to
\be
 q\,\Lambda^{\rm cl}_{LK}\langle B,Y\rangle .
\ee
Since the external $(L,K,Q)$ multiplet is omitted from the internal
mode space,  we have
\be
 \langle B,Y\rangle=0.
\ee
for every retained internal fluctuation.  Thus there is no
$O(qY)$ mixing between the external collective coordinate and
the internal Gaussian modes, and hence
we may  determine the one-loop self-energy of $q$ from
the Hessian evaluated on the displaced configuration
$\bar X=J+qB$.

The cubic interaction is schematically
\begin{align}
V_3[\overline X,Y]
&\sim \frac{M_P}{N^2}\Tr[\overline X,Y][Y,Y]
\nn\\
&=\frac{M_P}{N^2}\Tr[J,Y][Y,Y]
+\frac{qM_P}{N^2}\Tr[B,Y][Y,Y].
\label{displaced-cubic}
\end{align}
For a generic  ultraviolet internal fluctuation,
the adjoint derivative satisfies
\be
[J, Y] = \Order(N)Y.
\ee
As a result, like the round sphere, the first term gives the
maximal cubic scaling of $N^{-1}$. As for the second term,
because $B$ is unit normalized, it is
at most $\Order(qN^{-2})$  and is therefore subleading
for $q=\Order(1)$.
As for the quartic interaction, it is
\be
V_4 \sim
\frac{M_P}{N^2}
\Tr\!\left([Y,Y]^2\right).
\ee
Consequently, the maximal explicit $N$-scalings are
\be \label{N34}
\text{(UV cubic vertex)} \sim N^{-1},
\qquad
\text{(quartic vertex)} \sim N^{-2}.
\ee
Note that if the background commutator acts on a fixed-$L$ external line,
the cubic vertex is smaller. The order of $N$ in \eq{N34}  
is therefore an upper bound appropriate for UV-dominated internal lines.
Note also that for a  UV internal mode,
$\o_{\ell \sim N} =\Omega_0\sqrt{\L_\ell}=\Order(M_P)$ and so
we have
\be
\text{UV propagator}:N^0,
\quad
\text{frequency integral}:N^0,
\quad
\text{closed color face}:N.
\ee
We now organize the large-$N$ expansion using the
standard ’t Hooft double-line representation of the matrix Feynman diagrams
\cite{tHooft:1973alw,Brezin:1977sv}.
Consider a connected 1PI two-point ribbon graph with
$V_3$ cubic vertices, $V_4$ quartic vertices,
$I$ internal propagators, $F$ closed color-index faces, and $h$
frequency loops. Writing $V=V_3+V_4$, the topology of the ribbon graph gives
\be\label{t1}
 V -I +F = 2-2g -b,
\ee
where $g$ is the genus and $b$ is the number of boundaries.
For the present self-energy graphs,
the two external adjoint legs belong to a single boundary, so $b=1$.
The number of independent loop frequencies is
\be \label{t2}
h = I -V+1 ,
\ee
while counting half-edges gives
\be\label{t3}
3V_3 + 4V_4 = 2 I +2.
\ee
It follows that
\be
F= h -2g, \qquad h = \frac{V_3}{2} + V_4.
\ee
Using the explicit large-$N$ scaling \eq{N34} of the vertices,
together with one factor of $N$ for each closed color face,
the bosonic self-energy therefore scales as
\be
\Pi_{LK}^{B\,(h,g)}(0)\sim N^{-V_3-2V_4}N^F=\Order\!\left(N^{-h-2g}\right).
\ee
Thus each additional frequency loop suppresses the self-energy by one
power of $N$, while each handle gives the usual $N^{-2}$ nonplanar suppression.
The leading contribution is consequently the planar one-loop graph.
Note that this is an upper scaling. Angular-momentum selection rules
can suppress a coefficient, but under the stated assumptions no
diagram can carry a larger explicit power of $N$. As a result, the leading
bosonic
correction to the curvature comes from one-loop  planar  diagrams  and we have
\be\label{PB1}
\Pi_{LK}^B =\Order(1/N), \quad \implies \quad  \D \L^B_{LK} :=\Order(N).
\ee
We will confirm this result below with an explicit evaluation of the one-loop
determinant.

The large-$N$ counting of graphs containing fermion loops is more involved
than that of purely adjoint bosonic ribbon graphs, because the fermion
index structure is different and the fermion gap scales as $\D_F \sim N^{-1}$.
Nevertheless, the leading graph containing fermions is the one-loop Yukawa
bubble and 
\be \label{PF1}
\Pi_{LK}^F(0) =\Order(1/N^2), \quad \implies \quad  \D  \L^F_{LK} :=\Order(1)
\ee
is one order of $N$
smaller than \eq{PB1}. Here the $1/N^2$ dependence of the bubble graph
is a result of
a factor of  $1/N^4$ from
the two Yukawa vertices, a factor of $N$ from summing over
the $N$ flavors of fermions,
and a factor of $1/\D_F$ from the fermion loop integral
$
\int d\o\, G_+ (\o) G_-(\o) \sim \D_F^{-1},
$
where $\D_F:=E_+-E_- =a_2M_P/N \sim
M_P/N$ is the gap energy between the two bands of partons.

\subsubsection{Bosonic one-loop contribution to the self-energy}

The large-$N$ analysis above shows that the leading quantum correction
to the classical spectrum \eq{cs} comes from the planar bosonic
one-loop contribution and is of order $N$ in the curvature.  It does
not, however, determine its sign.  We therefore derive the explicit
one-loop contribution in terms of a sum over internal modes.  In
Sec.~3.2 we apply this result to the $(1,2)$ and $(2,3)$ modes, and in
Secs.~3.3 and 3.4 we use it to analyze the full spectrum.

For the purpose of extracting $C_{LK}$, we may choose a single
normalized magnetic component $B_a=B_a^{LKQ}$ and denote its amplitude
by $q$.  Rotational invariance guarantees that the resulting quadratic
coefficient is independent of $Q$; the full multiplet dependence is
restored at the end through the invariant
\be
|q_{LK}|^2:=\sum_{Q=-K}^{K}|q_{LKQ}|^2 .
\ee
As explained following \eq{XXbarY}, the
quadratic effective potential is determined by the displaced
Hessian
\be \label{cN}
 \cN(q)= \cN_0+q\cV_B+q^2\cW_B ,
\ee
where $\cN_0\equiv\cN_J$ is for the round fuzzy sphere, and
\be \label{VB}
(\cV_BY)_a :=-\Big(
[Y_b,[J_a,B_b]+[B_a,J_b]]
+[B_b,[Y_a,J_b]+[J_a,Y_b]]
+[J_b,[Y_a,B_b]+[B_a,Y_b]]
\Big)
\ee
and
\be \label{WB}
(\cW_BY)_a :=-\Big(
[Y_b,[B_a,B_b]] +[B_b,[Y_a,B_b]+[B_a,Y_b]]
\Big).
\ee
The quadratic action of the fluctuation is thus given by
\be
S[Y] = \frac{1}{2 a_0^2 M_P}\int dt \inner{Y}{\left[
    \partial_t^2 - \Omega_0^2\cN(q)\right] Y}.
\ee
Integrating over the internal fluctuations, we obtain the
one-loop effective action
\be\label{1loop}
\Gamma_B^{(1)}[q]=\frac12\Tr'\log D_B(q),
\ee
where $D_B(q)=-\partial_t^2+\Omega_0^2\cN(q)$ has the expansion
\be \label{DB}
D_B(q)=D_0+qD_1+q^2D_2,
\ee
\be \label{D012}
D_0:=-\partial_t^2+\Omega_0^2\cN_0,
\quad
D_1:=\Omega_0^2\cV_B,
\quad
D_2:=\Omega_0^2\cW_B.
\ee
Throughout the one-loop calculation, $\Tr'$ denotes the trace
over the strictly positive classical physical internal modes.
Thus the gauge branch, the classically non-positive $(1,2)$ and
$(2,3)$ multiplets, and the external collective multiplet
$(LKQ)_{Q= -K, \cdots, K}$ whose
curvature is being evaluated are omitted.
The classically non-positive $(1,2)$ and $(2,3)$ multiplets are treated
as collective low-energy directions rather than as Gaussian internal
fluctuations.  Since only finitely many low-angular-momentum modes are
removed, their omission does not affect the ultraviolet-dominated
leading large-$N$ coefficient considered below.

Introduce the propagator 
$G=D_0^{-1}$. The effective action \eq{1loop} can be expanded and gives
\be \label{G1}
\Gamma_B^{(1)}[q]
={}\Gamma_B^{(1)}[0]
+\frac q2\Tr'(GD_1) +q^2\left[
\frac12\Tr'(GD_2)
-\frac14\Tr'(GD_1GD_1)
\right]
+\Order(q^3).
\ee
To evaluate the effective action, we note that
the Hessian $\cN_0$ is diagonalized by the normalized matrix-valued
vector harmonics.  Denote the $n$th such mode by $Y_a^{(n)}$, so that
\be
 (\cN_0Y^{(n)})_a=k_nY_a^{(n)},
 \qquad
 \sum_a\Tr\!\left[(Y_a^{(n)})^\dagger Y_a^{(r)}\right]=\delta_{nr}.
 \label{modeinner}
\ee
Here $n =(\ell j m)$ is a collective label for the internal modes
  and $k_n = \Lcl_{\ell j}$.  
For compactness in notation,
let us denote the normalized vector harmonic $Y^{(n)}$ by
a mode-space ket $\ket n$ and the inner product by
\be
 \langle n|r\rangle
 :=\sum_a\Tr\!\left[(Y_a^{(n)})^\dagger Y_a^{(r)}\right],
 \qquad
 \bra{n}\cO \ket{r}
 :=\sum_a\Tr\!\left[(Y_a^{(n)})^\dagger(\cO Y^{(r)})_a\right],
 \label{brakettrace}
\ee
for any matrix $\cO_{ab}$. 
With this convention one may write $\cN_0\ket n=k_n\ket n$ and
define $\cV_{nr}=\bra{n}\cV \ket{r}$, $\cW_{nr}=\bra{n}\cW\ket{r}$. 
In frequency space,
\be
G_n(\nu)=\frac{1}{\nu^2+\Omega_0^2 k_n},
\qquad \Tr=\mathcal T\int_{-\infty}^{\infty}\frac{d\nu}{2\pi}\, \tr_{\rm modes},
\ee
where $\mathcal T$ is the Euclidean time interval and
$\Gamma_B^{(1)}=\mathcal T V_B^{(1)}$.
Since our purpose is to determine the self-energy, from this
point onward $\Delta V_B^{(1)}(q)$ denotes the part quadratic
in the external deformation.
As a result, we obtain
the one-loop correction to the bosonic potential 
\be \label{Vb1}
\D V_B^{(1)}(q)
=\frac{\Omega_0}{2}q_{LK}^2 C_{LK}(N),
\ee
where the coefficient
$C_{LK}(N)$ is given by
\be \label{CLK}
C_{LK}=C_{\mathrm{sg}}+C_{\mathrm{bub}},\qquad
C_{\mathrm{sg}}
:=\frac12\sum_n'\frac{\cW_{nn}}{\sqrt{k_n}},
\quad
C_{\mathrm{bub}}
:=-\frac14\sum_{n,r}'
\frac{|\cV_{nr}|^2}
{\sqrt{k_nk_r}
(\sqrt{k_n}+\sqrt{k_r})}.
\ee
To obtain these, we have 
used the frequency integrals
\be
\int\frac{\dd\nu}{2\pi}\,
\frac{1}{\nu^2+\Omega_0^2k_n}
=\frac{1}{2\Omega_0\sqrt{k_n}},\qquad
\int\frac{\dd\nu}{2\pi}\,
\frac{1}{(\nu^2+\Omega_0^2k_n)
(\nu^2+\Omega_0^2k_r)} =
\frac{1}{2\Omega_0^3\sqrt{k_nk_r}
(\sqrt{k_n}+\sqrt{k_r})} \nn
\ee
Note that diagrammatically,
the term $C_\mathrm{sg}$ is obtained from the $\cW$-$BBYY$ vertex by
contracting the two $Y$ legs, so it is called a seagull term. Similarly, the
term $C_\mathrm{bub}$ is obtained by contracting the $Y$'s of the
two  $\cV$-$BYY$ vertices, giving a bubble as a result.
Note also that the same result may be obtained
directly from the shifted zero point energy
\be
V_B^{(1)}(q)=\frac{\O_0}{2}\sum_n'\sqrt{k_n(q)}.
\ee
In fact if we denote
\be
k_n(q)=k_n+q a_n+q^2b_n+\cdots,
\ee
then it follows directly that
\be
\sum_n'\sqrt{k_n(q)}
=\sum_n'\sqrt{k_n}+ q A_{LK}(N) + q^2C_{LK}(N)+\Order(q^3),
\ee
with $C_{LK}$ given by \eq{CLK}. We remark that, due to rotational symmetry,
$A_{LK} =0$ for $K>0$ in general. The only exception is the scalar mode (1,0).
Since the present work concerns the
quadratic fluctuation spectrum, we restrict ourselves here to brief remarks
on the physical interpretation of this one-point function in the
discussion section, leaving a detailed analysis to a forthcoming publication.


Note that
the correction \eq{Vb1} in the potential may also be expressed as a correction
in the curvature
\be \label{dLC}
\D \L^{B}_{LK} = \frac{a_0 N}{2} C_{LK}(N).
\ee
Since we know from \eq{PB1} that
the correction is of order $N$ in the large $N$ limit, we conclude that
$C_{LK}$ is of order 1
\be \label{CLK1}
C_{LK} = \Order(1)
\ee
at large $N$. However the explicit result \eq{CLK} does not immediately
reveal the sign for
$C_{LK}$. We will analyze this further in the rest of this paper.

\subsection{Quantum stability of the $(1,2)$ and $(2,3)$ modes}
\label{qs-1223}

We know from \eq{cs} that the classical spectrum is positive except for
the (1,2) mode which is 
tachyonic with $\L_{12}^{\rm cl} =-2$, and the (2,3) mode which is
quadratically flat  with $\L_{23}^{\rm cl} =0$. It is
important to know what quantum corrections will do to these modes.
We now show with a direct computation that
\be
C_{12} (N)= \frac{4}{5}  - \frac{361}{10}\frac{1}{ N^2} + \Order(N^{-3}),
\ee
\be
  C_{23}(N)
 =\frac{104}{105}-\frac{135}{N^2}+\Order(N^{-3}).
 \ee
 The displayed asymptotic approximations change sign near $N \simeq 7$ and
 $N \simeq 12$, respectively.
 As a result,
the positive
$\Order(N)$ correction dominates the finite classical curvature $\L_{LK}^{\rm cl}$
and stabilizes these fluctuation modes.


\noindent\underline{Computation of $C_{12}(N)$}

To start, we note that in order to extract the coefficient $C_{LK}(N)$,
we can consider any $Q$ component of the perturbation
harmonics $B^{LKQ}$ 
since rotation invariance dictates the quantum potential to take
the form \eq{Vb1} and so every $Q$ component sees exactly the same $C_{LK}$.
For convenience, we will use the $Q=0$ component. It is technically preferable 
since it is Hermitian by itself
and enforces $m' =m$ in the internal perturbation matrices.

For $(L,K) =(1,2)$, we refer to \eq{An} and consider the following
perturbation 
\be \label{B-unnorm}
A_a :=S_{ab}J_b,
 \qquad
 S={\rm diag}\!\left(-\frac12,-\frac12,1\right) := {\rm diag}\, (s_1,s_2,s_3).
 \ee
 This has the operator norm
 \be \label{dN12}
 \sum_a\Tr(A_aA_a)= \frac{N(N^2-1)}{8}:=d_N
 \ee
 Thus the operator $B_a : = A_a/\sqrt{d_N}$
 is unit-normalized $\sum_a \Tr ( B_a^\dag B_a)=1$.
 Using \eq{VB} and \eq{WB}, we obtain
 \be
\cN(q) = \cN_0 + q \cV + q^2 \cW,
\ee
where
\be \label{cV}
\sqrt{d_N} \cV_{ab} =2Q_L\delta_{ab} -(s_a+s_b)L_bL_a +(s_a+s_b)[L_a,L_b],
\qquad Q_L:= \sum_cs_cL_c^2
 =\frac32L_3^2-\frac12\bm L^2,
 \ee
 and
 \be \label{cW}
d_N \cW_{ab} =R_L\delta_{ab}
 -s_as_bL_bL_a +s_as_b[L_a,L_b], \qquad R_L:= \sum_cs_c^2L_c^2
 =\frac14\bm L^2+\frac34L_3^2.
 \ee
We remark that in using the formula \eq{CLK}, the mode sum receives
 contributions from the two branches $j= \ell \pm 1$ as
 \bea
 j=\ell-1 &:& 
 \lambda_\ell^-=\ell(\ell+3),
 \qquad\qquad\;\; \ell=1,\ldots,N-1,\\
 j=\ell+1 &:& 
 \lambda_\ell^+=(\ell+1)(\ell-2),
 \qquad \ell=3,\ldots,N-1.
 \eea
 
To evaluate the matrix elements of $\cV$ and $\cW$, we use 
the Wigner--Eckart theorem. For the $Q=0$ component of a rank two tensor,
we have
\be \label{V-element}
 \langle\ell,j,m|\cV_0^{(2)}|\ell',j',m'\rangle
 =\delta_{\ell\ell'}\d_{m m'}(-1)^{j-m}
 \begin{pmatrix}
 j&2&j'\\
 -m&0&m
 \end{pmatrix}
 \langle\ell,j\Vert\cV^{(2)}\Vert\ell,j'\rangle.
\ee
The prefactor is easy to understand: for the operator \eq{cV}
that is built only from $L_a$,
it commutes with $\bm L^2$ and so  $\ell'=\ell$; as for $m'=m$,
it follows since  $Q=0$.
We note that a generic rank-two operator could mix $j=\ell-1$ and $j=\ell+1$.
However for the specific $\cV$ that  we have here, it turns out 
(see below for explicit computation) that
\be
 \langle\ell,\ell-1,m|\cV|\ell,\ell+1,m\rangle=0.
 \ee
Since the nonphysical gauge branch $j=\ell$ is omitted, $V$ is
diagonal in the retained physical basis $|\ell,j,m\rangle$.
Note that the diagonal $3j$ symbol takes the form
\be
 (-1)^{j-m}
 \begin{pmatrix}
 j&2&j\\
 -m&0&m
 \end{pmatrix}
 =\mathcal C_j\,[3m^2-j(j+1)],
\ee
where $\mathcal C_j$ is independent of $m$.
In fact, the polynomial factor on the RHS above
can be fixed without evaluating $\mathcal
C_j$: it must be even in $m$, quadratic, and traceless over the
complete spin-$j$ multiplet.  The unique such polynomial is
proportional to $3m^2-j(j+1)$.
Therefore, the diagonal elements $
v_{\ell jm}:=\langle\ell,j,m|\cV|\ell,j,m\rangle$ take the form
\be
v_{\ell jm}  =A_{\ell j}\,[3m^2-j(j+1)], \qquad j = \ell \pm 1
\ee
for some $A_{\ell j}$ independent of $m$.

To determine $v_{\ell jm} $, we introduce, at fixed magnetic number $m$,
the notation
\be
  \ket q_m:=Y_{\ell,m-q}e_q,
  \qquad q=+1,0,-1.
\ee
In the ordered basis $(\ket{+1}_m,\ket0_m,\ket{-1}_m)$, we can
use  $L_3Y_{\ell\mu}=\mu Y_{\ell\mu}$ and the ladder operator relations
and obtain  the matrix
$\widetilde{\cV}_{qq'} : = {}_m\langle q| \cV |q'\rangle_m$
\be
\sqrt{d_N} \; \widetilde{\cV} =
\begin{pmatrix}
-\frac{\cL}{2}+\frac52m^2-\frac72m+1
&\frac{\sqrt2}{4}(m-2)A_m
&-\frac12A_mB_m\\[2mm]
\frac{\sqrt2}{4}(m-2)A_m
&-\cL+m^2
&-\frac{\sqrt2}{4}(m+2)B_m\\[2mm]
-\frac12A_mB_m
&-\frac{\sqrt2}{4}(m+2)B_m
&-\frac{\cL}{2}+\frac52m^2+\frac72m+1
\end{pmatrix},
\ee
where
\be
  \cL:=\ell(\ell+1)
\ee
and
\be
  A_m:=\sqrt{(\ell+m)(\ell-m+1)},
  \qquad
  B_m:=\sqrt{(\ell-m)(\ell+m+1)}
\ee
are the standard orbital ladder operator coefficients. Similarly
\be
d_N \widetilde{\cW}=
\begin{pmatrix}
\frac{\cL}{8}+\frac78m^2-\frac{17}{8}m+\frac54
&\frac{\sqrt2}{4}(2-m)A_m
&\frac18A_mB_m\\[2mm]
\frac{\sqrt2}{4}(2-m)A_m
&\frac14(\cL-m^2)
&\frac{\sqrt2}{4}(m+2)B_m\\[2mm]
\frac18A_mB_m
&\frac{\sqrt2}{4}(m+2)B_m
&\frac{\cL}{8}+\frac78m^2+\frac{17}{8}m+\frac54
\end{pmatrix}.
\ee
Next, we record the Clebsch–Gordan coefficients.
Denote $ C_{q}^{\pm} := C_q^{(\ell\, (j= \ell \pm 1)\, m)}$ and let us package them as
a column vector.
For $j=\ell-1$, we have
\be
  \bm C^- :=
  \begin{pmatrix}
  C_+^-\\C_0^-\\C_-^-
  \end{pmatrix}
  =
  \begin{pmatrix}
  \displaystyle\sqrt{\frac{(\ell-m)(\ell-m+1)}{2\ell(2\ell+1)}}\\[3mm]
  \displaystyle-\sqrt{\frac{(\ell-m)(\ell+m)}{\ell(2\ell+1)}}\\[3mm]
  \displaystyle\sqrt{\frac{(\ell+m)(\ell+m+1)}{2\ell(2\ell+1)}}
  \end{pmatrix},
\ee
and for $j=\ell+1$,
\be
  \bm C^+ :=
  \begin{pmatrix}
  C_+^+\\C_0^+\\C_-^+
  \end{pmatrix}
  =
  \begin{pmatrix}
  \displaystyle\sqrt{\frac{(\ell+m)(\ell+m+1)}{2(\ell+1)(2\ell+1)}}\\[3mm]
  \displaystyle\sqrt{\frac{(\ell-m+1)(\ell+m+1)}{(\ell+1)(2\ell+1)}}\\[3mm]
  \displaystyle\sqrt{\frac{(\ell-m)(\ell-m+1)}{2(\ell+1)(2\ell+1)}}
  \end{pmatrix}.
\ee
Then we obtain the matrix elements of $\cV$ 
\be
  \bra{\ell,j_\sigma,m} \cV \ket{\ell,j_{\sigma'},m}
  = (\bm C^\sigma)^T\widetilde \cV\bm C^{\sigma'}
\ee
and an analogous formula for the matrix elements of $\cW$.
Using these, we  obtain immediately
 \be \label{v-ele-}
 v^-_{\ell m}
 =\frac{\ell+1}{ \ell \sqrt{d_N}}
 \left[3m^2-\ell(\ell-1)\right],
 \ee
 \be \label{v-ele+}
 v^+_{\ell m}
 =\frac{\ell}{(\ell+1) \sqrt{d_N}}
 \left[3m^2-(\ell+1)(\ell+2)\right].
 \ee
 and
\be \label{w-ele-}
 w^-_{\ell m}
 =\frac{
 9m^4 -3(\ell^2-9\ell-7)m^2
 +\ell(\ell-1)(\ell+1)(2\ell-5)
 }{4\ell(2\ell+1) d_N}
\ee
\be\label{w-ele+}
 w^+_{\ell m}
 =\frac{
 9m^4-3(\ell^2+11\ell+3)m^2
 +\ell(\ell+1)(\ell+2)(2\ell+7)
 }{4(\ell+1)(2\ell+1) d_N}.
\ee
We remark that a rank-two operator is kinematically
allowed to connect the two branches
$j_-=\ell-1$ and $j_+=\ell+1$ since $|j_+-j_-|=2$.
However, a direct computation shows that
the cross-branch matrix element vanishes,
$\bra{\ell,j_-,m} \cV \ket{\ell,j_{+},m} =0$. We note that this is not a
result of a selection rule, but  a special cancellation due to the specific
form of the vertex. One can also check that  the
analogous statement is not true for $\cW$. We find
\be
  \cW_{-+}
  =\frac{9(m^2+1)}{4d_N\sqrt{\ell(\ell+1)}(2\ell+1)}
  \sqrt{(\ell-m)(\ell-m+1)(\ell+m)(\ell+m+1)}
  \ee
  is generally nonzero. We note however that these mixed matrix elements
  are not needed since they do not  appear
  in the seagull term's mode sum \eq{CLK}. Only the diagonal elements
  $w_{\ell m}^\pm$ are required for the quadratic one-loop potential.

The results \eq{v-ele-}, \eq{v-ele+}, \eq{w-ele-}, \eq{w-ele+} 
can now be used to compute the coefficient $C_{LK}(N)$.
Summing the contributions
of the two branches in \eq{CLK}, we have
\be
 C_{LK}(N)
 =\sum_{\ell=1}^{N-1}c_\ell^-
 +\sum_{\ell=3}^{N-1}c_\ell^+
 \ee
 where
 \be
 c_\ell^\pm
 :=\frac{\beta_\ell^\pm}{2\sqrt{\lambda_\ell^\pm}}
 -\frac{\alpha_\ell^\pm}{8(\lambda_\ell^\pm)^{3/2}}
 \ee
 and 
\be
 \alpha_\ell^{\pm} :=\sum_m\left(v^\pm_{\ell m}\right)^2,
 \qquad
 \beta_\ell^{\pm} :=\sum_m w^\pm_{\ell m}
 \ee
 are obtained by summing the contributions over
 $m = -j, \cdots, j$ for $j = \ell \mp 1$.
 Evaluating them explicitly for the (1,2) mode,
 we have 
 \be
 \alpha_\ell^-
 =\frac{(\ell-1)(\ell+1)^2(2\ell-3)(2\ell-1)(2\ell+1)}{5\ell d_N},
\qquad
 \beta_\ell^-
 =\frac7{20 d_N}(\ell-1)(\ell+1)(2\ell-1),
\ee
\be
 \alpha_\ell^+
 =\frac{\ell^2(\ell+2)(2\ell+1)(2\ell+3)(2\ell+5)}{5(\ell+1)d_N},
 \qquad
 \beta_\ell^+
 =\frac7{20 d_N}\ell(\ell+2)(2\ell+3).
\ee
For large $N$, we can check from the form of $c_\ell^\pm$ given above that the
leading contribution is from the high angular momentum modes. For large $\ell$,
we obtain
\be
  d_Nc_\ell^-
  =\frac3{20}\ell^2+\frac3{10}\ell
  -\frac{349}{160}
  +\frac{1289}{160\ell}
  +\Order(\ell^{-2}),
\ee
\be
  d_Nc_\ell^+
  =\frac3{20}\ell^2
  -\frac{373}{160}
  -\frac{1289}{160\ell}
  +\Order(\ell^{-2}).
\ee
Note that the $1/\ell$ terms cancel exactly between branches,
and the combined remainder begins at $1/\ell^2$:
\be
  d_N(c_\ell^-+c_\ell^+)
  =\frac3{10}\ell^2+\frac3{10}\ell
  -\frac{361}{80}+\Order(\ell^{-2}).
\ee
As a result, 
there is no logarithmic term in the large-$N$ sum.
Now the $O(\ell^{-2})$ remainder has a convergent sum and contributes
only $\Order(1)$
to $d_NC_{12}$.  The isolated lower-branch shells $\ell=1,2$ also
contribute only $\Order(1)$.  Therefore
\begin{align}
  d_NC_{12}(N)
  &=\sum_{\ell=3}^{N-1}
  \left(\frac3{10}\ell^2+\frac3{10}\ell-\frac{361}{80}\right)
  +\Order(1)\\
  &=\frac1{10}N^3-\frac{369}{80}N+\Order(1).
\end{align}
Substituting \eq{dN12}, we obtain finally,
\be
 C_{12}(N)
 =\frac45-\frac{361}{10 N^2}+\Order(N^{-3}),
\ee
Note that the separately normalized contributions approach
\be
 C_{12, \rm sg}\longrightarrow\frac{28}{15},
 \qquad
 C_{12,\rm bub}\longrightarrow-\frac{16}{15},
\ee
leaving the limit of a positive total $4/5$.

\noindent\underline{Computation of $C_{23}(N)$}

For $(L,K) =(2,3)$, we obtain from \eq{Y230} the perturbation
\be \label{B23-unnorm}
 A_1=-\{J_1,J_3\},\qquad
 A_2=-\{J_2,J_3\},\qquad
 A_3=3J_3^2-\bm J^2.
 \ee
 To get its operator norm, we use
 \be
 \{J_-,J_3\}\ket{J,m}
 =(2m-1)\sqrt{\mathcal C-m(m-1)}\ket{J,m-1},
\ee
 where
 \be
 \mathcal C:=J(J+1)=\frac{N^2-1}{4},
\ee
then it is easy to obtain that
 \be \label{dN23}
 d_N := \sum_a\Tr(A_aA_a)= \frac{N(N^2-1)(N^2-4)}{12}.
 \ee
 Thus the operator $B_a := \frac{A_a}{\sqrt{d_N}}$ is unit-normalized
 $\sum_a\Tr(B_a^\dagger B_a)=1$.
 As a result,   we obtain the displaced Hessian
 \be
\cN(q) = \cN_0 + q \cV + q^2 \cW,
\ee
where $\cN_0$ is the same as before, and
\be \label{cV23}
(\cV Y)_a := (L_cE_c+E_cL_c)Y_a
 -(L_bE_a+E_bL_a)Y_b +\bigl([L_a,E_b]+[E_a,L_b]\bigr)Y_b,
 \ee
\be\label{cW23}
 (\cW Y)_a :=E_cE_cY_a-E_bE_aY_b+[E_a,E_b]Y_b
\ee
with
\be
 L_a :={\rm ad}_{J_a},
 \qquad
 E_a :={\rm ad}_{B_a}.
\ee
Unlike the (1,2) case, where $E_a \propto L_a$ preserves the orbital
shell $\ell$, here $E_a$ contains a commutator with a rank-two matrix
harmonic. The angular-momentum and commutator-parity selection rules
therefore allow it to connect an internal mode in shell $\ell$
to the adjacent shells $\ell \pm 1$.

To proceed, we express the Cartesian normalized vector harmonic
$B_a$ in terms of the 
{\it fuzzy spherical harmonics} $T_{LM}$ (see Appendix A for details).
Using  \eq{YTe}, we have
\be
 B_a=\sum_{M,q}\langle 2M;1q|30\rangle T_{2M}(e_q)_a.
\ee
Because $M+q=0$, only three terms occur. 
Let us denote $\gamma_q:=\langle 2,-q;1,q|30\rangle$, then
\be \label{BTe}
 B_a=\sum_{q=-1}^{1}\gamma_q T_{2,-q}(e_q)_a := \sum_q B_q (e_q)_a,
\ee
where we have denoted $B_q := \g_q T_{2,-q}$  as
the matrix coefficient of the spherical vector basis element $e_q$.
We also note that explicitly
\be
 \gamma_{\pm 1}=\frac1{\sqrt5},\qquad
 \gamma_0=\sqrt{\frac35}.
\ee

Next, let us determine the action of $E_q := {\rm ad}_{B_q}$ on the fuzzy
harmonics.
We start with the commutator algebra of normalized fuzzy spherical harmonics.
In general, for normalized fuzzy harmonics,
\begin{equation*}
 \Tr\!\left(T_{LM}^{\dagger}T_{L'M'}\right)
 =\delta_{LL'}\delta_{MM'},
 \qquad
 T_{LM}^{\dagger}=(-1)^M T_{L,-M}, \quad T_{00}=\bm 1_N/\sqrt N,
\end{equation*}
we have the product (in the Condon--Shortley convention) \cite{Chu:2001xi}, 
\begin{equation*}
 T_{LM}T_{\ell m}
 =\sum_{\ell'm'}
  P^{(N)}_{L\ell\ell'}
 \langle LM;\ell m|\ell'm'\rangle T_{\ell'm'}.
\end{equation*}
where
\begin{equation*}
  P^{(N)}_{L\ell\ell'}
 :=(-1)^{2s+\ell'}
 \sqrt{(2L+1)(2\ell+1)}
 \begin{Bmatrix}
  L&\ell&\ell'\\
  s&s&s
 \end{Bmatrix},
 \qquad s=\frac{N-1}{2}.
\end{equation*}
Reversing the two factors multiplies the coupled product by
$(-1)^{L+\ell-\ell'}$, we obtain the commutator algebra
\be
 [T_{LM},T_{\ell m}]
 =\sum_{\ell'm'} F^{(N)}_{L\ell\ell'}
 \langle LM;\ell m|\ell'm'\rangle T_{\ell'm'},
\ee
with the structure coefficient
\be \label{FLN}
  F^{(N)}_{L\ell\ell'}
:=\left[1-(-1)^{L+\ell-\ell'}\right]
 (-1)^{2s+\ell'}
 \sqrt{(2L+1)(2\ell+1)}
 \begin{Bmatrix}
  L&\ell&\ell'\\
  s&s&s
 \end{Bmatrix}.
\ee
For the present special case of $L=2$, the triangle rule gives
$|\ell-2|\le\ell'\le\ell+2$, while the factor in square brackets in \eq{FLN}
requires $2+\ell-\ell'$ to be odd, equivalently
$\ell+\ell'$ odd.  Consequently,
\be
\ell'=\ell\pm1.
 \label{ellselection}
\ee
we note that, unlike the (1,2) case where the bubble $C_{\rm bub}$ is given by a
sum preserving the orbital shell (i.e. $\ell = \ell'$),
the (2,3) bubble is intrinsically an
adjacent-shell sum: the diagonal $\ell = \ell'$ terms do not contribute.

Now with  $B_q=\gamma_qT_{2,-q}$ in the spherical basis,
we obtain the action of $E_q$ on the matrix harmonics
\be \label{ET}
E_q T_{\ell \m} =\sum_{\ell'\mu'}E^{(q)}_{\ell'\mu';\ell\mu}\,T_{\ell'\mu'},\qquad
\mbox{where}\quad
 E^{(q)}_{\ell'\mu';\ell\mu}
 :=\gamma_q F^{(N)}_{2\ell\ell'}
 C^{\ell'\mu'}_{2,-q;\ell,\mu}
\ee
 with  $E^{(q)}_{\ell'\mu';\ell\mu}$ nonvanishing only for
 $\ell'=\ell\pm1$ and $\mu'=\mu-q$.
Likewise, define $L^{(r)}_{\ell\mu';\ell\mu}$ by
\be\label{LT}
 L_rT_{\ell\mu}
 =\sum_{\mu'}L^{(r)}_{\ell\mu';\ell\mu}\,T_{\ell\mu'}, \qquad \mbox{where}\quad
 L^{(r)}_{\ell\mu';\ell\mu}
 :=(-1)^{\ell-\mu'}
 \begin{pmatrix}
 \ell&1&\ell\\ -\mu'&r&\mu
 \end{pmatrix}
 \sqrt{\ell(\ell+1)(2\ell+1)}.
\ee
The two actions \eq{ET} and \eq{LT} allow us to compute every matrix element of
$\cV$ and $\cW$ in terms of a finite number of computations involving the
Clebsch--Gordan coefficients.
 For later use, we record here the nonvanishing $F$-structure coefficients:
\be
 F_{\ell,+}:=F^{(N)}_{2,\ell,\ell+1},
 \qquad
 F_{\ell,-}:=F^{(N)}_{2,\ell,\ell-1},
\ee
we have
\be \label{F+-}
 |F_{\ell,+}|^2
 =\frac{5\ell(\ell+1)(\ell+2)}{2\ell+3}
 \frac{N^2-(\ell+1)^2}{d_N},\qquad
 |F_{\ell,-}|^2
 =\frac{5(\ell-1)\ell(\ell+1)}{2\ell-1}
 \frac{N^2-\ell^2}{d_N}.
\ee
To obtain these, we have used the Racah formula for the $6j$ symbol
\cite{Varshalovich:1988ifq}
\be
 \begin{Bmatrix}
 2&\ell&\ell+1\\ s&s&s
 \end{Bmatrix}^{\!2}
 =\frac{3\ell(\ell+1)(\ell+2)
 [N^2-(\ell+1)^2]}
 {(2\ell+1)(2\ell+3)N(N^2-1)(N^2-4)},
\ee
\be
 \begin{Bmatrix}
 2&\ell&\ell-1\\ s&s&s
 \end{Bmatrix}^{\!2}
 =\frac{3(\ell-1)\ell(\ell+1)
 [N^2-\ell^2]}
 {(2\ell-1)(2\ell+1)N(N^2-1)(N^2-4)}.
\ee
 
 Now we are ready to evaluate the matrix elements of $\cV$ and $\cW$. 
The Wigner--Eckart theorem for the $Q=0$ component of a rank three tensor gives
\be \label{WE3}
 \langle \ell',j',m'|\cV_0^{(3)}|\ell,j,m\rangle
 =\delta_{m'm}(-1)^{j'-m}
 \begin{pmatrix}
 j'&3&j\\[-1mm]
 -m&0&m
 \end{pmatrix}
 \langle \ell',j'\Vert\cV^{(3)}\Vert\ell,j\rangle .
\ee
The complete reduced matrix element is obtained by inserting the
elementary $E$ and $L$ matrix elements \eq{ET}, \eq{LT}  into
the three pieces of $\cV$:
\begin{align} \label{V123}
  \cV_1 &:=(L_cE_c+E_cL_c)\delta_{ab},\nn \\
  \cV_2 &:=-\,(L_bE_a+E_bL_a),\\
  \cV_3 &:=[L_a,E_b]+[E_a,L_b]. \nn
\end{align}
As we have explained above, the action of $E_a$ gives the selection rule
$\ell' = \ell \pm 1$ for
orbital shell. Let  us consider the upward transition
$\ell\to\ell+1$. As $j' = \ell' \pm 1$, we obtain $j' =\ell, \ell+2$.
Therefore there are four possible initial $(\ell, j)$ to final $(\ell+1, j')$
matrix  elements to consider:
\begin{align}
  -- &: (\ell,\ell-1)\longleftrightarrow(\ell+1,\ell),\\
  -+ &:(\ell,\ell-1)\longleftrightarrow(\ell+1,\ell+2),\\
  +- &: (\ell,\ell+1)\longleftrightarrow(\ell+1,\ell), \label{c+-}\\
  ++ &: (\ell,\ell+1)\longleftrightarrow(\ell+1,\ell+2).
\end{align}
We note that the downward transition
$\ell\to\ell-1$ gives precisely the reverse matrix elements. As the bubble
sum involves a sum over unordered adjacent-shell pairs:
$|\bra{r} \cV \ket{n}|^2$ and $\bra{r} \cV \ket{n}  = \bra{n} \cV \ket{r}^*$,
we only need to compute and include the matrix elements for the
upward channel $\ell' = \ell + 1$.
For the three nonzero upward-shell channels, let us define
\begin{align} 
 S_\ell^{--} &:=\sum_m
 \left|\mel{\ell+1,\ell,m}{\cV}{\ell,\ell-1,m}\right|^2,\label{SSS1}\\
 S_\ell^{-+}
 &:=\sum_m
 \left|\mel{\ell+1,\ell+2,m}{\cV}{\ell,\ell-1,m}\right|^2,\\ S_\ell^{++}
 &:=\sum_m \left|\mel{\ell+1,\ell+2,m}{\cV}{\ell,\ell+1,m}\right|^2.\label{SSS3}
\end{align}
The direct finite Clebsch–Gordan contraction of the three pieces of
the curvature vertex is worked out explicitly in 
Appendix \ref{S-23}. The result is
\be \label{Smm}
 S_\ell^{--}
 =\frac{8(\ell-1)(\ell+2)^2(2\ell-3)(2\ell-1)(2\ell+1)}
 {21(2\ell+3)d_N}
 \left[N^2-(\ell+1)^2\right],
\ee
\be\label{Spp}
 S_\ell^{++}
 =\frac{8\ell^2(\ell+3)(2\ell+3)(2\ell+5)(2\ell+7)}
 {21(2\ell+1)d_N}
 \left[N^2-(\ell+1)^2\right],
\ee
\be\label{Smp}
 S_\ell^{-+}
 =\frac{40(\ell+1)(2\ell-1)(2\ell+5)}
 {7(2\ell+1)(2\ell+3)d_N}
 \left[N^2-(\ell+1)^2\right].
\ee
These are the quantities that enter the bubble mode sum directly.
Note that the fourth kinematically allowed oriented channel $(+-)$ of \eq{c+-}
has a vanishing reduced matrix element.
This zero is not imposed by an angular-momentum selection rule.
Appendix \ref{S-23} shows explicitly that it results from
cancellation among the three pieces \eq{V123}  of the curvature vertex. 
Note also that the common  factor $N^2-(\ell+1)^2$ 
vanishes at $\ell=N-1$, as required by the fuzzy sphere harmonic cutoff.

For the seagull diagram only diagonal matrix elements of $\cW$ are needed.
Define the  magnetic sums
\be
 \beta_\ell^- :=
 \sum_{m=-(\ell-1)}^{\ell-1}
 \langle\ell,\ell-1,m|\cW|\ell,\ell-1,m\rangle,\qquad
 \beta_\ell^+ := 
 \sum_{m=-(\ell+1)}^{\ell+1}
 \langle\ell,\ell+1,m|\cW|\ell,\ell+1,m\rangle.
 \ee
We note that, since 
\be
 (\cW Y)_a=E_cE_cY_a-E_bE_aY_b+[E_a,E_b]Y_b
\ee
contains two adjoint actions of $E$,
an intermediate scalar harmonic may have orbital angular momentum
$\ell+1$ or $\ell-1$ before the second action of
$E$ returns it to the original shell.
Thus we can expect the magnetic trace  must be
organized into  a sum of two pieces proportional to $|F_{\ell,+}|^2$
and $|F_{\ell,-}|^2$. In fact carrying out the CG contractions, we obtain
\be
\beta_\ell^- =A_{-,+}(\ell)|F_{\ell,+}|^2
+A_{-,-}(\ell)|F_{\ell,-}|^2,
\ee
with
\be A_{-,+}(\ell) :=\frac{(2\ell-1)(2\ell+3)(25\ell-18)}
 {105\ell(2\ell+1)},\qquad
 A_{-,-}(\ell)
 :=\frac{2(2\ell-1)(4\ell^2-35\ell+12)}
 {105\ell(2\ell+1)},
 \ee
and
\be
 \beta_\ell^+
 =A_{+,+}(\ell)|F_{\ell,+}|^2
 +A_{+,-}(\ell)|F_{\ell,-}|^2,
 \ee
 with
 \be
 A_{+,+}(\ell)
 :=\frac{2(2\ell+3)(4\ell^2+43\ell+51)}
 {105(\ell+1)(2\ell+1)},\qquad
 A_{+,-}(\ell):=\frac{(2\ell-1)(2\ell+3)(25\ell+43)}
 {105(\ell+1)(2\ell+1)}.
 \ee
 Substituting the exact commutator coefficients and simplifying, we finally
 obtain
\be
 \beta_\ell^-=
 \frac{\ell+1}{21 d_N}
 \left[
 (29\ell^2-34\ell+12)N^2
 -29\ell^4-16\ell^3-51\ell^2+104\ell-36
 \right].
\ee
\be
 \beta_\ell^+=
 \frac{\ell}{21 d_N}
 \left[
 (29\ell^2+92\ell+75)N^2
 -29\ell^4-100\ell^3-177\ell^2-274\ell-204
 \right].
\ee

Using these, the  seagull term  is given by
\be
 C_{23,\rm sg}(N)
 =\sum_{\ell=1}^{N-1}\frac{\beta_\ell^-}{2\sqrt{\lambda_\ell^-}}
 +\sum_{\ell=3}^{N-1}\frac{\beta_\ell^+}{2\sqrt{\lambda_\ell^+}}
 \label{Csgexact}
\ee
and the bubble diagram term \eq{CLK} can be written as
\begin{align} \label{Cbubexact}
 C_{23,\rm bub}(N)
={}&-\sum_{\ell=2}^{N-1}
 \frac{S_\ell^{--}}
 {2\sqrt{\lambda_\ell^-\lambda_{\ell+1}^-}
 (\sqrt{\lambda_\ell^-}+\sqrt{\lambda_{\ell+1}^-})}
 -\sum_{\ell=2}^{N-1}
 \frac{S_\ell^{-+}}
 {2\sqrt{\lambda_\ell^-\lambda_{\ell+1}^+}
 (\sqrt{\lambda_\ell^-}+\sqrt{\lambda_{\ell+1}^+})}
 \nonumber\\
&\qquad-\sum_{\ell=3}^{N-1}
 \frac{S_\ell^{++}}
 {2\sqrt{\lambda_\ell^+\lambda_{\ell+1}^+}
 (\sqrt{\lambda_\ell^+}+\sqrt{\lambda_{\ell+1}^+})}.
\end{align}
Note that the factor of 1/4 has become 1/2 since the double sum in the
bubble term in \eq{CLK} has been converted into a sum over unordered
adjacent-shell pairs. We also note that the lower limits have simple origins.
The $--$ expression has an explicit $(\ell-1)$ factor and vanishes
at $\ell=1$. The $-+$ term at $\ell=1$ would connect to the
external $(2,3)$ collective mode and is therefore omitted.
The $++$ channel begins at $\ell=3$ because the upper physical branch
is included only from $\ell=3$.

To evaluate  the sum \eq{Csgexact} and \eq{Cbubexact},
we use the Euler–Maclaurin summation
formula. 
Set
\be
 x=\frac{\ell}{N},
 \qquad 0<x<1,
\ee
 the reciprocal frequencies expand as
 \be
 \frac1{\sqrt{\lambda_\ell^-}}
 =\frac1{Nx}
 \left(1-\frac{3}{2Nx}+\frac{27}{8N^2x^2}
 +\Order(N^{-3})\right),\quad
 \frac1{\sqrt{\lambda_\ell^+}}
 =\frac1{Nx}
 \left(1+\frac{1}{2Nx}+\frac{11}{8N^2x^2}
 +\Order(N^{-3})\right).
 \ee
 As a result, the lower- and upper-branch seagull summands
 \be
 c_{\rm sg,\ell}^-
 :=\frac{\beta_\ell^-}{2\sqrt{\lambda_\ell^-}},
 \qquad
 c_{\rm sg,\ell}^+
 :=\frac{\beta_\ell^+}{2\sqrt{\lambda_\ell^+}}
 \label{csgbranches}
 \ee
 has the expansion
 \begin{align}
 c_{\rm sg,\ell}^-
={}&\frac1N\frac{58}{7}(x^2-x^4)
 -\frac1{N^2}\frac{3x^3+97x}{7}
+\frac1{N^3}
 \frac{-1160x^4+381x^2+667}{28}
 +\Order(N^{-4}),
 \label{csgmexp}\\
 c_{\rm sg,\ell}^+
={}&\frac1N\frac{58}{7}(x^2-x^4)
 +\frac1{N^2}\frac{-229x^3+213x}{7}
 +\frac1{N^3}
 \frac{-1160x^4-975x^2+1287}{28}
 +\Order(N^{-4}).
 \label{csgpexp}
\end{align}
Combining the lower and upper seagull branches at the same
value of $\ell$ gives
\be
 c_{\rm sg,\ell}^-+c_{\rm sg,\ell}^+
 =\frac{f_{0,\rm sg}(x)}{N}
 +\frac{f_{1,\rm sg}(x)}{N^2}
 +\frac{f_{2,\rm sg}(x)}{N^3}
 +\Order(N^{-4}),
 \label{sgbulk}
\ee
where
\be \label{f-sg}
 f_{0,\rm sg}(x)=\frac{116}{7}(x^2-x^4),\quad
 f_{1,\rm sg}(x)=-\frac{232}{7}x^3+\frac{116}{7}x, \quad
 f_{2,\rm sg}(x)=-\frac{580}{7}x^4-\frac{297}{14}x^2+\frac{977}{14}.
 \ee

 As for the bubbles sum, consider first the bubble summands
 \be
 c_{\rm bub,\ell}^{--}
 :=-\frac{S_\ell^{--}}
 {2\sqrt{\lambda_\ell^-\lambda_{\ell+1}^-}
 (\sqrt{\lambda_\ell^-}+\sqrt{\lambda_{\ell+1}^-})}, \qquad
 c_{\rm bub,\ell}^{++}
 :=-\frac{S_\ell^{++}}
 {2\sqrt{\lambda_\ell^+\lambda_{\ell+1}^+}
 (\sqrt{\lambda_\ell^+}+\sqrt{\lambda_{\ell+1}^+})}.
 \ee
Using the identity $\frac1{\sqrt{ab}(\sqrt a+\sqrt b)}
 =\frac{a^{-1/2}-b^{-1/2}}{b-a}$,  and that for the adjacent shells,
$ \lambda_{\ell+1}^- -\lambda_\ell^-=2(\ell+2)$,
$ \lambda_{\ell+1}^+ -\lambda_\ell^+=2\ell$, we obtain the expansion
\begin{align}
 c_{\rm bub,\ell}^{--}
={}&\frac1N\frac{32}{7}x^2(x^2-1)
 -\frac1{N^2}\frac{64}{7}x(2x^2-3)
+\frac1{N^3}\frac4{7}
 (40x^4+55x^2-183)+\Order(N^{-4}),
 \label{cbubmmexp}\\
 c_{\rm bub,\ell}^{++}
={}&\frac1N\frac{32}{7}x^2(x^2-1)
 +\frac1{N^2}\frac{64}{7}x(6x^2-5)
+\frac1{N^3}\frac4{7}
 (40x^4+439x^2-311)+\Order(N^{-4}).
 \label{cbubppexp}
\end{align}
Their sum therefore has the form
\be
 c_{\rm bub,\ell}^{--}+c_{\rm bub,\ell}^{++}
 =\frac{f_{0,\rm bub}(x)}{N}
 +\frac{f_{1,\rm bub}(x)}{N^2}
 +\frac{f_{2,\rm bub}(x)}{N^3}
 +\Order(N^{-4}),
 \label{bubbulk}
\ee
where
\be \label{f-bub}
 f_{0,\rm bub}(x)=\frac{64}{7}(x^4-x^2),\quad
 f_{1,\rm bub}(x)=\frac{256}{7}x^3-\frac{128}{7}x, \quad
 f_{2,\rm bub}(x)=\frac{320}{7}x^4+\frac{1976}{7}x^2-\frac{1976}{7}.
 \ee
 For completeness, we note that
 the mixed term
 \be
 S_\ell^{-+}
 =\frac{40(\ell+1)(2\ell-1)(2\ell+5)}
 {7(2\ell+1)(2\ell+3)d_N}
 [N^2-(\ell+1)^2] 
 \ee
At fixed $x= \ell/N$, it is of order 
\be
 S_\ell^{-+}=\Order(N^{-2}).
\ee
Since  the oscillator denominator is of order $N^3$, it follows that
\be
 c_{\rm bub,\ell}^{-+}=\Order(N^{-5})
 \qquad (x\ \hbox{fixed}).
\ee
It  is therefore subleading and does
not enter the computation of the  leading or first subleading coefficient
of $C_{\rm bub}$. 

Now for a shell expansion
\be
 c_\ell=\frac{f_0(x)}N+\frac{f_1(x)}{N^2}
 +\frac{f_2(x)}{N^3}+\cdots,
 \qquad x=\frac\ell N,
\ee
the Euler–Maclaurin summation gives
the sum through order $N^{-2}$ as
\begin{align}
 \sum_{\ell=1}^{N-1}c_\ell
={}&\int_0^1f_0(x)\,dx\nonumber\\
&+\frac1N\left[
 \int_0^1f_1(x)\,dx
 -\frac{f_0(0)+f_0(1)}2
 \right]\nonumber\\
&+\frac1{N^2}\left[
 \int_0^1f_2(x)\,dx
 -\frac{f_1(0)+f_1(1)}2
 +\frac{f_0'(1)-f_0'(0)}{12}
 \right]
 +\Order(N^{-3}).
 \label{EM}
\end{align}
The finite differences between the exact lower limits in
Eqs.~\eqref{Csgexact} and \eqref{Cbubexact} affect only $\Order(N^{-3})$
and smaller terms.
Hence they do not alter the displayed coefficients through $N^{-2}$.
For the seagull functions \eq{f-sg}, we obtain
\be
 \int_0^1f_{0,\rm sg}\,dx=\frac{232}{105},
 \qquad
 \int_0^1f_{1,\rm sg}\,dx=0,
 \qquad
 f_{0,\rm sg}(0)=f_{0,\rm sg}(1)=0.
\ee
Substituting all terms in Eq.~\eqref{EM} yields
\be
 C_{23,\rm sg}(N)
 =\frac{232}{105}+\frac{155}{3N^2}+\Order(N^{-3}).
 \label{Csgasym}
\ee
Similarly, for the bubble functions \eq{f-bub}, we obtain
\be
 \int_0^1f_{0,\rm bub}\,dx=-\frac{128}{105},
 \qquad
 \int_0^1f_{1,\rm bub}\,dx=0,
 \qquad
 f_{0,\rm bub}(0)=f_{0,\rm bub}(1)=0,
\ee
and hence
\be
 C_{23,\rm bub}(N)
 =-\frac{128}{105}-\frac{560}{3N^2}+\Order(N^{-3}).
 \label{Cbubasym}
\ee
It is interesting to note that
the absence of a $1/N$ term holds separately for the seagull and bubble pieces.
Adding Eqs.~\eqref{Csgasym} and \eqref{Cbubasym}, we finally obtain
\be
 C_{23}(N)
 =\frac{104}{105}-\frac{135}{N^2}+\Order(N^{-3}).
 \label{C23asym}
\ee

\subsection{Quantum stability of the large-$N$ fuzzy sphere}
\label{qs-largeN}

We now turn from the two exceptional multiplets to the mechanism
controlling the full spectrum.
The mode sum analysis above demonstrates explicitly  the positivity
of $C_{LK}$ for the (1,2) and (2,3) modes. However this  method
must be applied  mode by mode and  becomes increasingly
complicated for general $(L,K)$. Moreover, 
the sign of $C_{LK}$ is not transparent because $C_{\rm sg}$ and $C_{\rm bub}$
have opposite signs. Therefore a more general method is needed.

In this subsection, we establish the large-$N$ stability of
every fixed physical mode. For this purpose, we introduce a
differential-form language which exposes an operator factorization
for the vertices entering the mode sum.
 The fluctuation $Y_a$ has one Cartesian index and will be regarded as
 a matrix-valued $1$-form.  An antisymmetric tensor $F_{ab}=-F_{ba}$
 is a matrix-valued $2$-form.  We denote the corresponding
 finite-dimensional Hilbert spaces by $\cH_1$ and $\cH_2$. Note that these are
 matrix-valued forms built from the adjoint derivatives;
 no continuum de Rham limit is required for the argument.

 For a general $X_a$, define
\be\label{dXY}
D_a:={\rm ad}_{X_a}=[X_a,\cdot].
\ee
The Hessian can be written
\be
(\cN_XY)_a=D_cD_cY_a-D_bD_aY_b+[D_a,D_b]Y_b-2Y_a.
\ee
On the $1$-form and $2$-form spaces introduce respectively
the inner product
\be
\langle Y,Z\rangle_1:=\sum_a\Tr(Y_a^\dagger Z_a),
\qquad
\langle F,G\rangle_2:=\frac12\sum_{a,b}\Tr(F_{ab}^\dagger G_{ab}).
\ee
Using \eq{dXY}, we introduce  the antisymmetrized adjoint derivative
\be
\mathsf d_X:\cH_1\longrightarrow\cH_2,
 \qquad
 (\mathsf d_XY)_{ab}:=D_aY_b-D_bY_a=2D_{[a}Y_{b]}.
\ee
Now since for any antisymmetric two-form $F_{ab}$, we have
\be
 \langle \mathsf d_XY,F\rangle_2=\sum_{a,b}\langle Y_b,D_aF_{ab}\rangle_1,
 \ee
this allows us to define the adjoint operator
\be
\mathsf d_X^\dag: \cH_2 \longrightarrow \cH_1,  \qquad
(\mathsf d_X^\dagger F)_b=D_aF_{ab}.
\ee
With this, we  note very interestingly
that the Hessian admits an exact decomposition of the
form
\be \label{NddR}
\cN_X = \mathsf d_X^\dag \mathsf d_X + \cR_X,
\ee
where
\be
(\cR_XY)_a :=[D_a,D_b]Y_b-2Y_a
\ee
is a remainder term. 
Note that the first term of \eq{NddR} is positive semi-definite
\be
 \langle Y,\mathsf d_X^\dagger \mathsf d_XY\rangle_1
 =\|\mathsf d_XY\|_2^2\ge0.
 \label{positiveSquare}
\ee
while the remainder is not sign-definite. So the study of the spectral 
positivity  reduces to determining how well
the remainder term can be controlled. 

To warm up, let us first consider the round fuzzy sphere $X_a = J_a$.
In this case, $D_a= L_a$ and we have, on an internal
matrix harmonic $Y_{\ell m}$ with orbital angular momentum $\ell$,
\be
 \mathsf d_0^\dagger \mathsf d_0 =\Order(\ell^2),\qquad
   [L_a,L_b] =i\epsilon_{abc}L_c=\Order(\ell),\qquad 
 -2=\Order(1)
\ee
and therefore
\be
 \cN_J
 =\mathsf d_0^\dagger \mathsf d_0
 \left[1+\Order\!\left(\frac1\ell\right)\right]
\ee
on the physical ultraviolet modes.
The exact eigenvalues verify this hierarchy:
\begin{align}
 k_\ell^-&
 =\ell(\ell+3),
 &&j=\ell-1,\label{minusSpectrum}\\
 k_\ell^+&
 =(\ell+1)(\ell -2),
 &&j=\ell+1.\label{plusSpectrum}
\end{align}
The common $\Order(\ell^2)$ term arises from the principal square,
while the branch splitting arises from the lower-derivative remainder.

Next, let us consider the displaced configuration \eq{XJB}. We write
\be
 D_a(q)=L_a+q\mathsf a_a,
 \qquad
 \mathsf a_a={\rm ad}_{B_a^{LKQ}}.
 \label{Ddeform}
\ee
Then the Hessian can be written as
\be \label{NddR2}
\cN(q) =  \cN_{\rm fact}(q)+ \cR(q),
\ee
where 
\be
 \cN_{\rm fact}(q): = \mathsf d(q)^\dagger \mathsf d(q)
 \label{Kfactq}
\ee
is the factorized part of the Hessian and $ \cR(q)$ defined by
\be
(\cR(q)Y)_a
=[D_a(q),D_b(q)]Y_b-2Y_a
\ee
is a non-factorized remainder.
Here in \eq{Kfactq}
\be \label{ddqb}
  \mathsf d(q) : = \mathsf d_0+q\, \mathsf b, \quad \mbox{where}\quad
 (\mathsf bY)_{ab}
 := \mathsf a_aY_b- \mathsf a_bY_a.
\ee

\noindent\underline{Factorized part $\cN_{\rm fact}(q)$}

It is instructive to consider the seagull-bubble expansion.
Substitute \eq{ddqb} and expand \eq{Kfactq} in $q$,
we obtain the $q$ and $q^2$ terms:
\be
 \cV_{\rm fact}
 =\mathsf d_0^\dagger \mathsf b+ \mathsf b^\dagger \mathsf d_0, \qquad
 \cW_{\rm fact}
 =\mathsf  b^\dagger \mathsf b \ge0.
 \ee
This relation between $\cV_{\rm fact}$ and $\cW_{\rm fact}$
allows us to derive 
\be \label{CCC}
C_{LK}^{\rm fact} \geq  0 
\ee
for the factorized part of the Hessian. To see this,
recall  from \eq{dXY} that $d_0$ maps a matrix-valued 1-form to an
antisymmetric matrix-valued $2-$form, 
\be
\mathsf d_0:\mathcal H_1
 \longrightarrow \mathcal H_{2}.
 \ee
The operator
$\cN_{0, \rm fact} =\mathsf d_0^\dag \mathsf d_0$
is an operator on the Hilbert space $\cH_1$
and we can consider its eigenvalue equation:
\be
\mathsf d_0^\dag \mathsf d_0 \ket{r} = \o^2_r \ket{r}.
\ee
We emphasize that $\omega_r^2$ is an eigenvalue of the
factorized operator and should not be confused with the full
Hessian eigenvalue $k_r$.
Here $\ket{r}$ is one of the unperturbed physical vector modes $Y_a^{(r)}$,
schematically $\ket{r} = \ket{\ell, j, m}, j = \ell \pm 1$
after omitting the gauge branch. Acting with $\mathsf d_0$ on $\ket{r}$ produces
a normalized antisymmetric tensor,
\be
 |u_r\rangle=\frac{1}{\omega_r}\mathsf d_0|r\rangle.
\ee
In components, $|u_r\rangle$ is represented by
\be
 (u_r)_{ab}=\frac{1}{\omega_r}
 \bigl(L_aY_b^{(r)}-L_bY_a^{(r)}\bigr).
\ee
We emphasize that the right singular vectors $|r\rangle$ of
$\mathsf d_0$ belong to the
physical vector-fluctuation space $\cH_1$, whereas the left singular vectors
$|u_r\rangle$ belong to the antisymmetric two-form space $\cH_2$.
They live in different spaces and cannot be identified.
Using these, we obtain the matrix elements
\be \label{Wfact}
 (\cV_{\rm fact})_{nr}
 =\omega_n b_{nr}+\omega_r b_{rn}^*,
\qquad 
(W_{\rm fact})_{rr}=\mel r{b^\dagger b}r=\|\mathsf b\ket r\|_2^2
\geq\sum_n|b_{nr}|^2,
\ee
where  $b_{nr } :=\langle u_n|\mathsf b |r\rangle$.
Note that the inequality holds
because in general the left singular vectors
$\{\ket{u_n}\}$ span only $\operatorname{im} \mathsf d_0\subset\cH_2$.
Equality holds if $\mathsf b\ket r$ has no component orthogonal to
$\operatorname{im} \mathsf d_0$. We can now use the
Cauchy–Schwarz inequality
\be
 |\omega_nx+\omega_ry|^2
 \le(\omega_n+\omega_r)
 \left(\omega_n|x|^2+\omega_r|y|^2\right)
 \label{weightedCS}
\ee
with equality iff $x=y$. 
With  $x=b_{nr}$ and $y=b_{rn}^*$, we obtain
\be
 \frac{|(\cV_{\rm fact})_{nr}|^2}
 {\omega_n\omega_r(\omega_n+\omega_r)}
 \le
 \frac{|b_{nr}|^2}{\omega_r}
 +\frac{|b_{rn}|^2}{\omega_n}.
 \label{pairBound}
\ee
 Summing over $n,r$ gives
\begin{align}
 |C_{LK,\rm bub}^{\rm fact}|
 &\le\frac12\sum_r
 \frac{\sum_n|b_{nr}|^2}{\omega_r}
 \leq\frac12\sum_r
 \frac{(\cW_{\rm fact})_{rr}}{\omega_r}
 =C_{LK,\rm sg}^{\rm fact}.
 \label{seagullBubbleBound}
\end{align}
Therefore
\be\label{Cfact-positive}
C_{LK}^{\rm fact} \geq 0.
\ee
Thus factorization gives a rigorous non-negativity result.
We note that the Cauchy-Schwarz inequality is saturated if
\be
 b_{nr}=b_{rn}^* ,
 \label{schwarz-saturation}
\ee
while the inequality \eq{Wfact} is saturated if $b|r\rangle$ lies entirely in
the image of $d_0$.
For a generic physical deformation there is no reason for these
conditions to hold simultaneously for all internal modes and so
$C^{\rm fact}_{LK}>0$.

Interestingly,
the rotational invariant $(L,K)=(1,0)$ scale mode is exceptional.
For this mode both inequalities are saturated identically mode by mode
and hence $C^{\rm fact}_{10}=0$.
See  Appendix \ref{scalingmode}. There, we also show that, by including
the remainder part, 
\be
 C_{10}(N)
 =
 -\frac{16}{N^2}
 +O(N^{-3}).
\ee
Therefore the one-loop corrected curvature is
\be
 \Lambda^{\rm eff}_{10} =
 4-\frac{8a_0}{N}
 +O(N^{-2})
\ee
and is positive $ \Lambda^{\rm eff}_{10}>0$
for sufficiently large $N$.  Thus the scale mode is stable.

The general non-negativity result \eq{Cfact-positive}
can also be obtained by considering the zero point energy. In
fact, if we ignored the contribution from the remainder part,
the zero-point energy for the displaced configuration is given by
\be
 E_{\rm fact}(q)
 =\frac{\O_0}{2}\Tr'\sqrt{\mathsf d(q)^\dagger \mathsf d(q)}
 =\frac{\O_0}{2}\| \mathsf d(q)\|_{1}'.
 \ee
Here $\Tr'$ denotes the same fixed internal mode space used
throughout the one-loop calculation: the strictly positive
classical physical internal modes, with the gauge branch, the
classically non-positive $(1,2)$ and $(2,3)$ multiplets, and
the external collective multiplet omitted.  The norm
$\|\cdot\|_1$ denotes the corresponding nuclear norm, 
the sum of singular values on
this fixed physical mode space.  Note that the convexity of the norm gives
$ E_{\rm fact}((1-t)q_1+tq_2)
 \le(1-t)E_{\rm fact}(q_1)+tE_{\rm fact}(q_2)$.
Taking $q_1=-h$, $q_2=h$, and $t=1/2$ gives the second order differences
$ E_{\rm fact}(h)+E_{\rm fact}(-h)-2E_{\rm fact}(0)\ge0$ and hence
$E_{\rm fact}''(0) \geq 0$.
With the convention
\be
 E_{\rm fact}(q)
 =E_{\rm fact}(0)+E_{\rm fact}^\prime(0)q+\frac{\O_0}{2} C_{LK}^{\rm fact}q^2+\cdots,
\ee
we obtain
\be
C_{LK}^{\rm fact}\geq 0.
 \label{CfactPositive}
\ee
Now since the factorized curvature is a sum of non-negative terms,
a vanishing large-$N$ limit would require saturation of the
factorization inequality at leading order throughout the whole ultraviolet
sector.
The $(1,0)$ analysis shows explicitly that saturation requires a
special structural mechanism.  For the scale mode this mechanism is
the exact proportionality $\mathsf b\propto \mathsf d_0$.
No analogous uniform-rescaling
identity exists for all the other modes with $K>0$.  We therefore expect the
factorized ultraviolet contribution of a non-saturating physical mode
to approach a strictly positive limit,
\be
 C^{\rm fact}_{LK}(N)
 \longrightarrow
 c^{\rm fact}_{LK}>0 .
\ee

We remark that, in applying the convexity argument, we keep the
round-sphere physical projector $P_{\rm phys}$
fixed as $q$ is varied. The exact gauge
projector for the displaced configuration is $q$-dependent, but for an
external perturbation of angular momentum $L$ acting on an internal mode
of angular momentum $\ell$, its deviation from the round-sphere projector
is lower derivative,
\be
\d P_{\rm phys} = \Order(\frac{L}{\ell}).
\ee
Since the one-loop coefficient is dominated by
ultraviolet internal modes
with $\ell \sim N$, this gives an $\Order(L/N)$
correction and does not affect the leading factorized UV  curvature
considered here.

\noindent\underline{Non-factorized remainder $\cR (q)$}

We can similarly expand $\cR(q)$ in $q$. We obtain
\be \label{RRVW}
 \cR(q)=\cR_0+q \cV_{\rm rem}+q^2 \cW_{\rm rem},
\ee
with
\be
 (\cV_{\rm rem}Y)_a
 =\bigl([L_a,\mathsf a_b]+[\mathsf a_a,L_b]\bigr)Y_b,\qquad
 (\cW_{\rm rem}Y)_a
 =[\mathsf a_a,\mathsf a_b]Y_b.
\ee
Note that these vertices are not controlled by the Cauchy-Schwarz inequality.
Their size determines how far the positivity theorem extends
beyond fixed external $L$. Let us consider
an external harmonic with angular momentum $L$,
\be
 [L_a,\mathsf a_b]
 ={\rm ad}_{[J_a,B_b]}=\Order(L)\,\mathsf a.
\ee
On an internal mode with angular momentum $\ell$,
\be
 \cV_{\rm fact}\sim \mathsf d_0^\dagger\mathsf b+\mathsf b^\dagger \mathsf d_0
 =\Order(\ell)\, \mathsf a.
\ee
Hence
\be
 \frac{\cV_{\rm rem}}{\cV_{\rm fact}}
 =\Order\!\left(\frac{L}{\ell}\right).
\ee
For dominant internal modes $\ell\sim N$, this is $\Order(L/N)$. Therefore
corrections from the remainder $\cV_{\rm rem}$ are suppressed as $L/N$.
As for the $\cW$-vertex, we have
\be
 \cW_{\rm fact}\sim\mathsf a^\dagger\mathsf a, \quad
\mbox{versus}\quad
 \cW_{\rm rem}\sim[\mathsf a,\mathsf a]
 ={\rm ad}_{[B,B]}.
\ee
For a rank-$L$ fuzzy harmonic, the star-product expansion
\cite{Steinacker:2024unq}
gives
\be
 [B^{(L)},B^{(L)}]
 =\frac{i}{N}\{B^{(L)},B^{(L)}\}_{\rm P.B.}
 +\cdots.
\ee
Since the Poisson bracket differentiates each factor once,
\be
 \{B^{(L)},B^{(L)}\}_{\rm P.B.}
 =\Order(L^2)\,B^{(L)}B^{(L)},
\ee
we have the estimate
\be \label{quadraticRatio}
 \frac{\cW_{\rm rem}}{\cW_{\rm fact}}
 =\Order\!\left(\frac{L^2}{N}\right).
 \ee
 This estimate is parametrically small in the regime
$L\ll\sqrt{N}$.  At the crossover scale $L\sim\sqrt{N}$ it becomes
$O(1)$, and the crossover region must therefore be treated separately,
as we will do next in Sec.~3.4.

Now, since the  mode sum is dominated by internal modes with high
angular momentum $\ell = x N$, $0 < x \leq 1$, 
the complete coefficient is given by
\be \label{C-w-error}
 C_{LK}(N)
 =C_{LK}^{\rm fact}(N)+\Delta C_{LK}(N),
 \qquad
 C_{LK}^{\rm fact}(N) \geq 0, \qquad \Delta C_{LK}
 \sim\Order\!\left(\frac1N,\frac{L}{N},\frac{L^2}{N}\right),
\ee
where the correction $\Delta C_{LK}$
is due to the presence of $\cR$
of \eq{RRVW}, and  receives contributions from
the $1/N$ corrections of the fuzzy sphere remainder $\cR_0$, and the
$L/N$ and $L^2/N$ corrections of the  remainder terms
$\cV_{\rm rem}$ and $\cW_{\rm rem}$.
For a fixed physical mode $(L,K)$, the
non-factorized contribution vanishes as $N\to \infty$.
On the other hand,
for every fixed non-saturating physical mode, the factorized contribution
approaches a strictly positive limit. Hence
\be
C_{LK}(N) = c_{LK}^{\rm fact}+ o(1) >0
  \ee
 for sufficiently large $N$. The exceptional
$(1,0)$ scale mode saturates the factorization inequality and 
 has  $C_{10} = \Order(N^{-2}) <0$. Nevertheless its one-loop corrected
 curvature is
 positive $ \Leff_{10} = 4 + \Order(N^{-1}) >0$.
Together with the explicit stabilization of the classically tachyonic
(1,2) mode and the classically marginal (2,3) mode, this establishes
pointwise large-$N$ stability of every fixed physical mode.
In the next subsection, we establish uniform positivity
over the complete physical spectrum at sufficiently large finite rank $N$.

\subsection{Quantum stability at sufficiently large finite $N$}
\label{qs-finiteN}

The preceding subsection establishes \emph{pointwise} large-$N$
stability: every fixed physical mode has positive total
one-loop-corrected 
curvature for sufficiently large $N$. For generic non-saturating modes,
this follows from eventual positivity of $C_{LK}$,
while the exceptional (1,0) scale mode retains positive corrected curvature
by classical-curvature dominance even though $C_{10} <0$.
The stronger statement needed for the finite-rank fuzzy sphere is
uniform positivity of the complete physical spectrum.  Namely,
we want to show that there exists a single finite $N_{\rm crit}$ such that
\be
    \Lambda^{\rm eff}_{LK}(N)>0
    \label{uniform-goal}
\ee
for every physical mode present at rank $N$ whenever $N>N_{\rm crit}$.

Before presenting the analytic argument, let us first record the
direct numerical evidence supporting this conclusion. We have
performed exhaustive evaluations of the complete physical spectrum at
$N=30,50$ and $100$.  At each of these ranks, every physical multiplet on both
branches $K=L\pm1$ has positive one-loop corrected curvature,
as summarized in Table~\ref{tab:finiteN-complete-scan}.
We note that, as discussed in Appendix \ref{scalingmode}, only
for the (1,0) mode does the one-loop
effective potential also contain a linear term, whose physical
implication will be treated separately. Apart from this,
the  question is therefore not whether the
complete finite-rank curvature
stability can occur, but rather why this stability
persists at sufficiently large $N$. 

\begin{table}[t]
\centering
\caption{Exhaustive finite-rank scans of the static bosonic one-loop
  spectrum.  The number of physical multiplets includes both branches $K=L\pm1$.}
\label{tab:finiteN-complete-scan}
\begin{tabular}{c|c|c|c}
\hline\hline
$N$ & physical multiplets & mode of minimum & $\min\Lambda^{\rm eff}_{LK}$\\
\hline
30  & 58  & $(1,0)$ & $3.723513$\\
50  & 98  & $(1,0)$ & $3.833258$\\
100 & 198 & $(1,0)$ & $3.916392$\\
\hline\hline
\end{tabular}
\end{table}
For either physical branch,
\be
\Lambda^{\rm cl}_{LK}=L^2+O(L),
\qquad K=L\pm1,
\label{classical-largeL}
\ee
while the leading bosonic one-loop correction is
\be
\Delta\Lambda^B_{LK}=\frac{a_0N}{2}\,C_{L,K}(N).
\label{quantum-shift}
\ee
The competition between the two terms is therefore governed by
how $L$ grows with $N$. 
To proceed, it is convenient to use the
scaled variables to parameterize the external momentum by
\be
x := \frac{L}{ N}.
\ee
We have, for any fixed $x>0$,
\be
\Lambda^{\rm cl}_{LK}=x^2N^2+O(N), \qquad \Delta\Lambda_{LK}=O(N),
\label{fixedx-orders}
\ee
so that
\be
\frac{|\Delta\Lambda_{LK}|}{\Lambda^{\rm
    cl}_{LK}}=O\!\left(\frac1N\right).
\label{fixedx-dominance}
\ee
Thus the linearly scaling sector $L\sim N$ becomes increasingly
classical. Any potentially delicate family must arise from the boundary
region $x \sim 0$.

Let us introduce the scaled variable
\be
y := \frac{\ell}{ N}
\ee
to parameterize the  internal angular momenta. 
We note that the one-loop mode sum is dominated by
ultraviolet internal angular momenta modes with $y=O(1)$. After
simplification of angular momentum coefficients,
the one-loop mode sum can be written as a sum over momentum shells as
\be
 C_L^{\pm}(N)=\frac{1}{N}\sum_{\ell}f^{\pm}(N,x,y).
 \label{shell-representation}
\ee
Here we have denoted $C_L^{\pm}(N):=C_{L,L\pm1}(N)$ for the two branches. A
normalization factor $1/N$ is adopted since there
are $N$ shells in the sum. 
For a regular planar limit one expects
\be
 f^{\pm}(N,x,y)
 =f_0^{\pm}(x,y)
 +\frac{1}{N}f_1^{\pm}(x,y)
 +\frac{1}{N^2}f_2^{\pm}(x,y)+\cdots,
 \label{kernel-expansion}
\ee
so that
\be\label{fixedx-expansion}
 C_L^{\pm}(N)
 =F_0^{\pm}(x)
 +\frac{1}{N}F_1^{\pm}(x)
 +\frac{1}{N^2}F_2^{\pm}(x)+\cdots,
 \qquad x=\frac{L}{N},
 \ee
 where $F_0^\pm(x)=\int_0^1dy\,f_0^\pm(x,y)$, and
 the higher $F_n^\pm(x)$ denote the corresponding coefficients
in the large-$N$ expansion. 
%
To investigate the behaviour of  $C_L^{\pm}(N)$ near the endpoint, we consider
the ``endpoint'' value,
\be \label{bulk-endpoint}
 F_{\pm}:=F_0^{\pm}(0).
\ee
A convenient way to approach this limit is to take
\be
 L=N^{\alpha},\qquad 0<\alpha<\frac12.
 \label{alpha-path}
\ee
Along this trajectory, the non-factorizing remainder vanishes, so
the factorized determinant controls the endpoint. Assuming that
the $K>0$ endpoint family does not asymptotically saturate the
factorization inequality, its factorized ultraviolet contribution
has a strictly positive limit.  Hence
\be
 F^\pm>0.
\ee
This non-saturation assumption is supported by the crossover
numerics presented in Appendix \ref{app:numerics}.

We next analyze the crossover sector $L = c \sqrt{N}$.
Substituting $x = c/\sqrt{N}$ to \eq{fixedx-expansion},
and assuming that the functions $F_n^\pm(x)$ are smooth near $x=0$, 
we obtain immediately
a large $N$ expansion 
for $C^\pm_L$ in powers of $N^{-1/2}$:
\be\label{CAN}
C_L^{\pm}(N) = F_\pm + \frac{a_\pm c}{\sqrt{N}} +\cdots,
\ee
where
\be
a_\pm := F_0^{\pm}{}'(0). 
\ee
It is important to note that $F_\pm$, as defined by \eq{bulk-endpoint},
is independent of $c$. 
The expansion \eq{CAN} implies that 
\be
\lim_{\substack{N\to\infty\\L/\sqrt N\to c}} C_L^{\pm}(N) = F_\pm >0
\qquad\mbox{for every fixed finite $c$}.
\ee
Hence for every fixed finite $c$, there exists an $N_c$ such that
\be \label{C+}
C_L^\pm(N) >0, \qquad N > N_c, \qquad L= c \sqrt{N}.
\ee
Note that 
$N_c$ depends on $c$ in general.
If the coefficient $a_\pm$ in
\eq{CAN} is negative and nonzero, then the leading estimate for $N_c$ is
\be
 N_c=O(c^2).
\label{Nc-c2}
\ee
Assuming that the expansion is uniform on every compact interval
$0\le c\le c_0$,
\eq{fixedx-expansion} gives a single finite rank $\widebar{N}_{c_0}$
\be
\widebar{N}_{c_0} := \sup_{0<c<c_0} N_c
\ee
above which $C_L^{\pm}>0$ for the whole sector of modes.

The dependence of $N_c$ on $c$ means that \eq{C+} is a
pointwise statement in $c$, and not a uniform statement over arbitrarily large
$c$.  This however causes no difficulty
because the classical curvature  takes over when
$c=L/\sqrt N$ becomes large.  In fact, in
the large-$L$, small-$x$ sector, the
same regularity assumption underlying \eq{fixedx-expansion} keeps the
one-loop coefficient bounded, say $|C_{LK}(N)|\le M$.  Then
\be
 \frac{|\Delta\Lambda_{LK}|}{\Lambda^{\rm cl}_{LK}}
 \lesssim \frac{a_0M}{2}\frac{N}{L^2}
 =\frac{a_0M}{2c^2} = \Order\left(\frac{1}{c^2}\right).
\label{largec-dominance}
\ee
As a result, there exists a certain critical $c_{\rm crit}$ such that
the curvature is classically dominated for $c\geq c_{\rm crit}$.
Thus $C_{LK}$
need not remain positive for arbitrarily large $c$, positivity of the
total curvature  instead follows from the classical dominance.

Summarizing, the complete large-$N$ picture is 
simple. For finite $c=L/\sqrt{N}$, the crossover expansion gives
$C^\pm_L(N) \to F_\pm >0$ , so these modes are stable for sufficiently large $N$.
When $c$ becomes large, positivity of  $C^\pm_L$ is
no longer required: the classical curvature grows as $L^2 =c^2 N$,
whereas the one-loop correction is $\Order(N)$,
and is therefore suppressed relative to the classical term by $\Order(c^{-2})$.
Under the stated smoothness and non-saturation assumptions, these two
mechanisms, together with the fixed-$L$ result of Sec.~3.3, cover the
complete physical spectrum and establish stability for sufficiently
large finite $N$.

Numerical tests of the smoothness and non-saturation assumptions
entering this argument,
together with exhaustive finite-rank checks of the spectrum, are presented
in Appendix \ref{app:numerics}.

\section{Discussion}
In this paper we have established local curvature stability of the
fuzzy-sphere horizon at sufficiently large finite $N$, under the
smoothness and non-saturation assumptions described in Sec.~3.4 and
supported numerically in Appendix~D. Large-$N$ counting
shows that the leading quantum correction is governed by the planar
bosonic one-loop contribution, and we have shown that the resulting
quantum fluctuation spectrum is positive for sufficiently large $N$.

In the analysis, an intermediate angular momentum scale $L \sim
\sqrt{N}$ emerges naturally. We note that at low angular momentum $L
\ll \sqrt{N}$, a fluctuation has a wavelength much larger than the
microscopic noncommutative scale and therefore probes the fuzzy sphere
essentially as a smooth classical geometry. At the opposite extreme $L
\sim N$, the fluctuation approaches the angular-momentum cutoff of the
matrix algebra and becomes sensitive to the microscopic structure of
the fuzzy sphere.  Interestingly, the crossover occurs at the
mesoscopic scale $L \sim \sqrt{N}$, where classical and quantum
contributions to the fluctuation curvature become parametrically
comparable and neither can be neglected.  One may wonder whether 
other interesting physics is associated with this  intermediate
regime of angular
momentum. It is intriguing that, when the monopole configuration
\be
A^{(g)} : = X^{(N)} - X^{(N-g)}
\ee
appearing in the
tunneling process \cite{Chu:2026wyh} is decomposed into angular momentum modes
$B^{LKQ}$, the expansion is found to be likewise dominated by modes with
$L \sim \sqrt{N}$. The appearance of the same mesoscopic scale in two
rather different problems suggests that this classical--quantum, or
IR-UV crossover regime may contain further nontrivial physics
of the quantum black hole.
An appropriate effective description of the horizon in this regime is
neither a purely classical metric sphere nor the full microscopic matrix model,
but rather a metric sphere equipped with the fuzzy-sphere star-product
algebra of functions \cite{Steinacker:2024unq},
whose leading noncommutative correction is governed by
the Poisson bracket.
It would be interesting to understand this regime more systematically
and to determine whether the mesoscopic scale $L\sim\sqrt{N}$ plays a
role in other distinctive aspects of the quantum dynamics of the black hole.

It is interesting to note that the (1,0) scaling mode is exceptional
in that a linear term in its effective potential is generated quantum
mechanically. If the fuzzy sphere were considered in isolation, this
tadpole would imply that the round configuration is no longer a
stationary point of the quantum effective action and
would lead to a shift of its radius. 
This situation is reminiscent of the treatment of tadpoles in string theory.
In the Fischler--Susskind mechanism, a nonvanishing NS--NS
tadpole signals that the perturbative
expansion of string theory has been performed about a background
which is not a solution of the quantum-corrected equations of motion;
the tadpole is then absorbed into the backreacted background
\cite{Fischler:1986ci,Fischler:1986tb}. 
Now a black-hole
horizon is not an isolated system: it separates an interior
spacetime from an exterior spacetime and must ultimately be coupled
consistently to the surrounding gravitational degrees of freedom. In
this setting, the (1,0) tadpole need not represent a pathology of the
fuzzy-sphere background, but may instead signal that the horizon
sector by itself does not constitute a closed quantum background. Its
implications for the gravitational sector of the horizon membrane will
be investigated in a forthcoming publication \cite{prep}.

The present result also invites comparison with the classical
stability of black holes in general relativity. Black hole uniqueness theorems
\cite{Israel:1967wq, Carter:1971zc,Carter:1973rla,Robinson:1975bv}
characterize the possible stationary end states,
while linear perturbation theory determines whether a black hole is
dynamically stable against small disturbances. Linear perturbations
about the Schwarzschild solution are described by the
odd-parity Regge--Wheeler and the even-parity Zerilli equations
\cite{Regge:1957td,Zerilli:1970se}.
Although the Regge--Wheeler and Zerilli potentials are different,
the two parity sectors are related via the Chandrasekhar transformation
\cite{Chandrasekhar:1975nkd}, resulting in an identical spectrum of 
quasinormal mode frequencies
\cite{Chandrasekhar:1975zza}. In the eikonal regime
$\ell \gg 1$, the real part grows linearly with angular momentum as
\cite{Cardoso:2008bp}
\be
{\rm Re}\; \o_{\ell n}^{\rm QNM} \simeq \frac{\ell+\frac{1}{2}}{3 \sqrt{3} M},
  \ee
  while the imaginary part is controlled by the Lyapunov  exponent
  of the unstable null orbit. As a result, parametrically,
  \be \label{oGR}
     {\rm Re}\; \o_{\ell n}^{\rm QNM} \sim \frac{\ell}{R}.
     \ee
     The fuzzy-sphere horizon exhibits a suggestive similarity.  Its
physical fluctuation branches are $K=L\pm1$, while the $K=L$ branch
consists of gauge modes.  Under spatial inversion, the scalar fuzzy
harmonics have parity
\be
P T_{LM}=(-1)^L T_{LM}.
\ee
Since the vector index of $B_a^{LKQ}$ transforms as a polar vector, it
follows that
\be
P B_a^{LKQ}=(-1)^{L+1}B_a^{LKQ}.
\ee
Since the two physical branches have $K=L\pm1$, their parity may be
written as $(-1)^K$, whereas the gauge branch $K=L$ has parity
$(-1)^{K+1}$.  This suggests the schematic parity correspondence
\be
\begin{array}{rcl}
K=L\pm1
&:&
B_a^{LKQ}
\ \longleftrightarrow\
\Psi^{\rm even}_{KQ}
\qquad\text{(Zerilli)},\\[2mm]
K=L
&:&
B_a^{LLQ}
\ \longleftrightarrow\
\Psi^{\rm odd}_{LQ}
\qquad\text{(Regge--Wheeler)} .
\end{array}
\label{parity-correspondence}
\ee
The second line is only schematic at this stage, since the $K=L$
mode is pure gauge for the isolated fuzzy sphere.
There is also a suggestive correspondence in the dispersion relation.
At large angular momentum, the classical fuzzy-sphere frequencies
satisfy
\be
\omega_{LK}^{\rm hor}\sim \frac{K}{R},
\label{fuzzy-dispersion}
\ee
while the real part of the Schwarzschild quasinormal-mode frequency
obeys
\be
{\rm Re}\,\omega^{\rm QNM}_{Kn}\sim \frac{K}{R}.
\label{qnm-dispersion}
\ee
Thus the horizon modes and the bulk gravitational modes exhibit the
same large-$K$ angular-momentum scaling.
The correspondences above should not, however, be interpreted as a
direct identification.  The frequency \eqref{fuzzy-dispersion}
describes a normal mode of the isolated fuzzy-sphere horizon, whereas
the black-hole quasinormal-mode frequencies describe resonances of the
coupled horizon--bulk gravitational system.  It would therefore be
interesting to determine how, after coupling and locking \cite{Chu:2026vzj} the
fuzzy-sphere degrees of freedom to bulk gravity,
the horizon modes identified in this paper couple
to the Regge--Wheeler and Zerilli modes in the bulk.  In particular,
it is interesting to ask whether the $K=L$ gauge modes become physical
as relative horizon--bulk degrees of freedom after locking.

\appendix

\addtocontents{toc}{\protect\setcounter{tocdepth}{1}}

\section{Spherical harmonic tensor 
  and fuzzy spherical harmonics}
\label{harmonics}

Rotational symmetry can be described in two complementary languages
\cite{Thorne:1980rmp,Varshalovich:1988ifq}:
\begin{enumerate}[label=(\roman*)]
\item In the \emph{spherical basis},
  irreducible states are labeled by angular momentum and its
  magnetic quantum number,
\be
 |L,M\rangle,\qquad M=-L,-L+1,\ldots,L.
\ee
This basis is adapted to the commuting operators $\bm L^2$
and $L_3$ and is therefore the natural basis for manipulations involving
angular momentum coupling using the Clebsch--Gordan coefficients.

\item In the \emph{Cartesian tensor basis}, a spin-$L$ object is
  represented by a rank-$L$ tensor carrying indices $a_i=1,2,3$.
  This language is natural when the fundamental variables are Cartesian
  vectors such as coordinates $x_a$, matrices $J_a$, or vector
  fluctuations $\delta X_a$.
\end{enumerate}
Introduced by Thorne \cite{Thorne:1980rmp},
the {\it rank-$L$ spherical harmonic tensor}
\be
 \cE^{(LM)}_{a_1\cdots a_L}
\ee
is the change-of-basis tensor connecting these two descriptions.  Its
contraction with a Cartesian rank-$L$ tensor extracts the component
with definite $(L,M)$, just like a polarization vector extracts a
spherical component of an ordinary vector.
Starting
with the Cartesian orthonormal basis $\bm e_1,\bm e_2,\bm e_3$, we define
\be
 \bm e_{+1}=-\frac{\bm e_1+i\bm e_2}{\sqrt2},
 \qquad
 \bm e_0=\bm e_3,
 \qquad
 \bm e_{-1}=\frac{\bm e_1-i\bm e_2}{\sqrt2}.
 \label{spherical-vector-basis}
\ee
We take $L$ copies of the spin-one representation
and project them onto the maximal total spin $L$.  The resulting tensor is
\be\label{E-multicg}
 \cE^{(LM)}_{a_1\cdots a_L}
 :=
 \sum_{q_1,\ldots,q_L=-1}^{1}
 \langle 1q_1;\ldots;1q_L\mid LM\rangle
 (\bm e_{q_1})_{a_1}\cdots(\bm e_{q_L})_{a_L}.
\ee
Note that $\cE^{(LM)}_{a_1\cdots a_L}$ is obviously symmetric. It is 
traceless since
$\d_{a_1 a_2} (\bm e_{q_1})_{a_1} (\bm e_{q_2})_{a_2}= (-1)^{q_1}\d_{q_1, -q_2}$. 
An equivalent recursive definition is
\be
 \cE^{(LM)}_{a_1\cdots a_L}
 =
 \sum_{M',q}
 \langle L-1,M';1q\mid LM\rangle
 \cE^{(L-1,M')}_{a_1\cdots a_{L-1}}
 (\bm e_q)_{a_L},
 \label{E-recursive}
\ee
with the initial condition
\be
 \cE^{(1M)}_a=(\bm e_M)_a,
 \qquad
 \cE^{(00)}=1.
 \label{E-initial}
\ee
The spherical harmonic tensors satisfy the following properties:
\begin{align}
\mbox{Complete symmetry:} & \qquad \qquad 
\cE^{(LM)}_{a_1\cdots a_L}
 =\cE^{(LM)}_{(a_1\cdots a_L)}, \\
\mbox{Tracelessness:}&\qquad \qquad 
 \delta_{a_1a_2}\cE^{(LM)}_{a_1a_2a_3\cdots a_L}=0, \\
\mbox{Orthonormality:}&\qquad \qquad 
 \cE^{(LM)*}_{a_1\cdots a_L}
 \cE^{(LM')}_{a_1\cdots a_L}
 =\delta_{MM'},\\
\mbox{Complex conjugation:}&\qquad \qquad 
 \cE^{(LM)*}_{a_1\cdots a_L}
 =(-1)^M\cE^{(L,-M)}_{a_1\cdots a_L},\\
\mbox{Rotation law:} &\qquad \qquad 
 R_{a_1b_1}\cdots R_{a_Lb_L}
 \cE^{(LM)}_{b_1\cdots b_L}
 =
 \sum_{M'=-L}^{L}
 \cE^{(LM')}_{a_1\cdots a_L}
 D^{(L)}_{M'M}(R).
\end{align}
The $2L+1$ tensors $\cE^{(LM)}$ form an orthonormal basis of
the rank-$L$ symmetric-traceless space.  The completeness relation is
\be
 \sum_{M=-L}^{L}
 \cE^{(LM)}_{a_1\cdots a_L}
 \cE^{(LM)*}_{b_1\cdots b_L}
 =
 \Delta^{(L)}_{a_1\cdots a_L;b_1\cdots b_L},
 \label{E-completeness}
\ee
where $\Delta^{(L)}$ is the identity operator on symmetric-traceless
rank-$L$ tensors. For example,
\be
\Delta^{(1)}_{a;b}=\delta_{ab}, \qquad
 \Delta^{(2)}_{ab;cd}
 =\frac12(\delta_{ac}\delta_{bd}+\delta_{ad}\delta_{bc})
 -\frac13\delta_{ab}\delta_{cd}.
 \ee

 As mentioned, the spherical harmonic tensor is the bridge between
 the spherical components and Cartesian components of a symmetric traceless
 tensor.
 A symmetric-traceless tensor $A_{a_1\cdots a_L}$ has the spherical components
\be
 A_{LM}=\sum_{a_1, \cdots, a_L}\cE^{(LM)*}_{a_1\cdots a_L}A_{a_1\cdots a_L}.
 \label{spherical-component}
\ee
The inverse relation is
\be
 A_{a_1\cdots a_L}
 =\sum_{M=-L}^{L}A_{LM}\cE^{(LM)}_{a_1\cdots a_L}.
 \label{inverse-spherical-component}
\ee
 The tensor is  harmonic since in general
 given any completely symmetric traceless
 rank-$L$ tensor $A_{a_1\cdots a_L}$, the homogeneous polynomial
\be
 P_L(\bm x) :=A_{a_1\cdots a_L}x_{a_1}\cdots x_{a_L}.
 \label{homogeneous-polynomial}
\ee
is harmonic
\be
 \nabla^2 P_L =0.
 \ee
 If we restrict to the unit sphere
\be
 P_L(\bm x)=r^L p_L(\bm n), \qquad \bm x=r\bm n,
 \quad \bm n^2=1,
\ee
then $p_L$  is a spherical harmonic
of angular momentum $L$ since
\be
 \Delta_{S^2}p_L=-L(L+1)p_L.
 \label{sphere-eigenvalue}
\ee
If we take the spherical harmonic tensor \eq{E-multicg}
and include the normalization
properly, we obtain 
the normalized scalar spherical harmonics
\be
Y_{LM}(\bm n)
 =
 \sqrt{\frac{(2L+1)!!}{4\pi L!}}\,
 \cE^{(LM)}_{a_1\cdots a_L}
 n_{a_1}\cdots n_{a_L}.
 \ee

 The spherical harmonic tensor allows one to introduce the fuzzy
 sphere matrix harmonics immediately \cite{Madore:1991bw,Ramgoolam:2001zx}. 
 Let $J_a$ be the generators of the spin-$j$ irreducible representation of
 $SU(2)$
\be
 [J_a,J_b]=i\epsilon_{abc}J_c,
 \qquad
 J_aJ_a=j(j+1)\id,
 \qquad
 N=2j+1.
\ee
Then we can construct a {\it rank-$L$ fuzzy (matrix) spherical harmonics}
$T_{LM}$ as
\be\label{T-from-E}
 T_{LM}
 :=\mathcal N_{Lj}\,
 \cE^{(LM)}_{a_1\cdots a_L}
\cJ^{(L)}_{a_1\cdots a_L},\qquad L=0,1,\ldots,N-1
\ee
where 
\be
\cJ^{(L)}_{a_1\cdots a_L}
 ={\rm STF}\!\bigl[J_{(a_1}\cdots J_{a_L)}\bigr].
 \ee
 is the degree-$L$ symmetric-traceless operator form from $J_a$'s.
 Here STF (symmetric trace-free) stands for the operation of
symmetrizing over all indices and subtracting out every possible trace
\cite{Thorne:1980rmp}.
$\mathcal N_{Lj}$ is a normalization factor. For example, it can be chosen
such that
\be
 \Tr(T_{LM}^\dagger T_{L'M'})
 =\delta_{LL'}\delta_{MM'}.
\ee
Group theoretically, the matrix harmonics satisfy
\begin{align}
& [J_a,[J_a,T_{LM}]]=L(L+1)T_{LM},\\
& [J_\pm, T_{LM}] = \sqrt{(L\mp M)(L \pm M +1)} T_{L, M\pm1}, \qquad
 [J_3,T_{LM}]=M T_{LM},\\
& T_{LM}^\dagger=(-1)^M T_{L,-M}.
\end{align}

For our work, we are interested in a
matrix-valued vector fluctuation $\delta X_a$. Decomposed, it carries
an orbital angular momentum $L$ and a spin 1 from its Cartesian index $a$.
A vector matrix harmonic with total angular momentum $(K,Q)$ thus takes
the form 
\be \label{YTe}
\bigl(\mathcal Y^{LKQ}\bigr)_a
 =
 \sum_{M,q}
 \langle LM;1q\mid KQ\rangle
 T_{LM}(\bm e_q)_a.
 \ee
 Substituting \eq{T-from-E},  we obtain
 a representation directly in terms of $J_a$
 \be
 \bigl(\mathcal Y^{LKQ}\bigr)_a
 =\mathcal N_{Lj}
 \cE^{(LKQ)}_{a;a_1\cdots a_L}
 \cJ^{(L)}_{a_1\cdots a_L}, \qquad 
 \cE^{(LKQ)}_{a;a_1\cdots a_L}
 := \sum_{M,q}
 \langle LM;1q\mid KQ\rangle
 (\bm e_q)_a
 \cE^{(LM)}_{a_1\cdots a_L}.
\ee
 This equation shows how the product of a spin-$L$ STF tensor and a vector
 becomes a tensor of total angular momentum $K$.

 We note that for the branch $K =L+1$,  we have the maximal-spin coupling
 $ L\otimes1\longrightarrow L+1$ and so
\be
 \cE^{(L,L+1,Q)}_{a;b_1\cdots b_L}
 =
 \cE^{(L+1,Q)}_{a b_1\cdots b_L},
 \label{maximal-branch-identity}
\ee
where the RHS is the harmonic spherical tensor \eq{E-multicg}. Hence it is
symmetric and traceless.
For example, for the $(L,K)=(1,2)$ mode, we have
 \be \label{Y12Q}
 \bigl(\mathcal Y^{1,2,Q}\bigr)_a
 =\cN_1
 \cE^{(2Q)}_{a;b}J_{b},
\ee
where $\cE^{(2Q)}_{a;b}$ is traceless and symmetric. 
For the specific case of $Q=0$, we have
\be
\cE^{(20)}_{a;b} = \frac{1}{\sqrt{6}}(3 \d_{a3} \d_{b3}-\d_{ab}).
\ee
This gives
\be \label{Y120}
\cY^{1,2,0}_a =  \frac{\cN_1}{\sqrt{6}}(-J_1, -J_2, 2J_3).
\ee
Similarly, for the $(L,K)=(2,3)$ mode, we have
\be
 \bigl(\mathcal Y^{2,3,Q}\bigr)_a
 = \cN_2
 \cE^{(3Q)}_{a;bc}Q_{bc}, \qquad 
 Q_{ab}=\frac12\{J_a,J_b\}-\frac13\delta_{ab}J^2.
\ee
In particular, we have  for $Q=0$
\be
\cE^{(30)}_{a;bc} = \frac{1}{\sqrt{10}}[5 \d_{a3}\d_{b3}\d_{c3} -\d_{ab}\d_{c3}
  -\d_{bc}\d_{a3} -\d_{ca}\d_{b3}].
\ee
Since $Q_{ab}$ is symmetric and traceless, we obtain
\be
\cE^{(30)}_{a;bc} Q_{bc} = \frac{1}{\sqrt{10}}(5 \d_{a3} Q_{33} -2 Q_{a3}).
\ee
This gives
\be \label{Y230}
\cY^{2,3,0}_1 = -\frac{\cN_2}{\sqrt{10}}\{J_1,J_3\}, \qquad
\cY^{2,3,0}_2 = -\frac{\cN_2}{\sqrt{10}}\{J_2,J_3\}, \qquad
\cY^{2,3,0}_3 = \frac{\cN_2}{\sqrt{10}}( 3 J_3^2 -J^2).
\ee 

\section{Explicit derivation of \texorpdfstring{
    $S_\ell^{++}$, $S_\ell^{--}$, $S_\ell^{-+}$,
    $S_\ell^{+-}$}
  {S++---++-}
  for the (2,3) mode}
\label{S-23}

We are interested in the computation of the matrix element
\be \label{MV}
M_{j'j}(m):=\mel{\ell+1,j',m}{\cV}{\ell,j,m}
\ee
that appears in the magnetic bubble-sum \eq{SSS1}-\eq{SSS3}
for the vertex created by the (2,3)-mode.
We recall \eq{BTe} that the $Q=0$ component of the external $(2,3)$-mode reads
\be
 B_a=\sum_{q=-1}^{1}\gamma_q T_{2,-q}(e_q)_a,
 \qquad
  \gamma_{+1}=\gamma_{-1}=\frac1{\sqrt5},
 \quad
 \gamma_0=\sqrt{\frac35}.
 \ee
 Let us consider the  upward orbital transition $\ell\to\ell+1$. The three
 pieces \eq{V123} of $\cV$ take the explicit form
\begin{align}
 (\mathcal V_{1})_{ab}
 &=\delta_{ab}\sum_c
 \left(L_c^{\rm out}\mathcal E_c+\mathcal E_cL_c^{\rm in}\right),
 \label{appV1}\\
 (\mathcal V_{2})_{ab}
 &=-\left(L_b^{\rm out}\mathcal E_a+\mathcal E_bL_a^{\rm in}\right),
 \label{appV2}\\
 (\mathcal V_{3})_{ab}
 &=L_a^{\rm out}\mathcal E_b-\mathcal E_bL_a^{\rm in}
   +\mathcal E_aL_b^{\rm in}-L_b^{\rm out}\mathcal E_a.
 \label{appV3}
\end{align}
Here, in order to facilitate computation,  we have introduced
the  angular momentum matrices acting on the incoming and outgoing spaces as
\be
 L_a^{\rm in}:=L_a^{(\ell)},
 \qquad
 L_a^{\rm out}:=L_a^{(\ell+1)},
\ee
where $L_a^{(r)}$ is for an angular momentum $r$,
\be
 L_3^{(r)}T_{r\mu}=\mu T_{r\mu},
 \qquad
 L_\pm^{(r)}T_{r\mu}
 =\sqrt{(r\mp\mu)(r\pm\mu+1)}\,T_{r,\mu\pm1}.
 \label{appLaction}
\ee
As for the operator $\cE_a$, it is defined to be the $(\ell+1)$-component of
$E_a T_{\ell\mu}$,
\be
 \mathcal E_a T_{\ell\mu}
 :=\left.E_aT_{\ell\mu}\right|_{\ell+1}
 =F_{\ell,+}\sum_{q=-1}^{1}
 \gamma_q(e_q)_a
 C^{\ell+1,\mu-q}_{2,-q;\ell,\mu}
 T_{\ell+1,\mu-q}.
 \label{appEup}
\ee
Equivalently, we can define the matrix with the elements
\be
 (\mathcal E_a)_{\nu\mu}
 =F_{\ell,+}\sum_q\gamma_q(e_q)_a
 C^{\ell+1,\nu}_{2,-q;\ell,\mu},
 \qquad \nu=\mu-q.
 \label{appEmatrix}
\ee
The Wigner--Eckart theorem gives
\be
 M_{j'j}(m):=\mel{\ell+1,j',m}{V}{\ell,j,m}
 =(-1)^{j'-m}
 \begin{pmatrix}j'&3&j\\-m&0&m\end{pmatrix}
 R_{j'j}^{(\ell)}.
 \label{appWEfull}
\ee
For the computation below, it is convenient to
introduce the partial  reduced matrix element
\be
R^{[s]}_{j'j}(\ell):=\langle \ell+1,j'\Vert \cV_s \Vert\ell,j\rangle,
\qquad s=1,2,3.
\ee
By definition, their sum gives the reduced matrix element of the full vertex
\be
 R_{j'j}^{(\ell)}
 =\sum_{s=1}^{3}R^{[s]}_{j'j}(\ell).
\ee
It is interesting to note that
\be
 \sum_m\begin{pmatrix}j'&3&j\\-m&0&m\end{pmatrix}^{\!2}=\frac17
 \label{app3jsum}
\ee
and so the desired bubble sum
\be \label{S17R}
 S_{j'j}^{(\ell)}=\frac17\,|R_{j'j}^{(\ell)}|^2
 \ee
 can be determined once $R$ is computed.

 We will now determine
 $R_{j'j}^{(\ell)}$ and $R_{j'j}^{[s]}$ with a direct evaluation of the matrix
 element
 \be\label{RMRM}
 R_{j'j}^{(\ell)}
 =\frac{M_{j'j}(m)}
 {(-1)^{j'-m}
   \begin{pmatrix}j'&3&j\\-m&0&m\end{pmatrix}},\qquad
 R_{j'j}^{[s]} =\frac{M^{[s]}_{j'j}(m)}
 {(-1)^{j'-m}
   \begin{pmatrix}j'&3&j\\-m&0&m\end{pmatrix}}.
\ee
Before that, we note that the $R$'s are  independent of $m$, therefore we can
pick any convenient choice of $m$ to perform the computation. 
Let's start with the determination of $R_{--}$. 
Take
\be
 j=\ell-1,\qquad j'=\ell,
\ee
and for convenience,
\be
 m=\ell-1.
\ee
For the incoming vector harmonic write
\be
 (Y^{\ell,\ell-1,\ell-1})_a
 =\sum_{p=-1}^{1}c_p\,T_{\ell,\ell-1-p}(e_p)_a,
\ee
where the three Clebsch--Gordan coefficients are
\be
 c_{+1}=\frac1{\sqrt{\ell(2\ell+1)}},\qquad
 c_0=-\sqrt{\frac{2\ell-1}{\ell(2\ell+1)}},\qquad
 c_{-1}=\sqrt{\frac{2\ell-1}{2\ell+1}}.
 \label{appmmcin}
\ee
For the outgoing harmonic on the $\ell+1$ shell,
\be
 (Y^{\ell+1,\ell,\ell-1})_a
 =\sum_{p=-1}^{1}d_p\,T_{\ell+1,\ell-1-p}(e_p)_a,
\ee
with
\be
 d_{+1}=\sqrt{\frac{3}{(\ell+1)(2\ell+3)}},\qquad
 d_0=-2\sqrt{\frac{\ell}{(\ell+1)(2\ell+3)}},\qquad
 d_{-1}=\sqrt{\frac{\ell(2\ell+1)}{(\ell+1)(2\ell+3)}}.
 \label{appmmcout}
\ee
Substituting Eqs.~\eqref{appmmcin}--\eqref{appmmcout}, the
$E$-action \eqref{appEup}, and the $L$-action
\eqref{appLaction} into the definition \eq{MV},
performing the finite sum over $p,p',q=-1,0,1$,
and noticing that the common factor $F_{\ell,+}$ (see \eq{F+-})
can be singled out because $\mathcal E_a$
appears linearly in each of the partial vertex $\cV_s$, we obtain 
\be
 \frac{M^{[1]}_{--}(\ell-1)}{F_{\ell,+}}=-(2\ell+5)\mathcal A_\ell,
 \qquad
 \frac{M^{[2]}_{--}(\ell-1)}{F_{\ell,+}}=2\mathcal A_\ell,
 \qquad
 \frac{M^{[3]}_{--}(\ell-1)}{F_{\ell,+}}=2\mathcal A_\ell,
 \label{appmmMpieces}
\ee
where
\be
 \mathcal A_\ell
 :=\frac{4(\ell-1)(2\ell-3)}{\sqrt5\,\ell(\ell+1)(2\ell+1)}
 \sqrt{\frac{(\ell+2)(2\ell-1)}{2\ell+3}}
 \label{appAell}
\ee
and
\be
|F_{\ell,+}|^2
 =\frac{5\ell(\ell+1)(\ell+2)}{2\ell+3}
 \frac{N^2-(\ell+1)^2}{d_N}.
 \ee
Summing, we obtain
\be 
\frac{M_{--}(\ell-1)}{F_{\ell,+}}
 =-(2\ell+1)\mathcal A_\ell.
 \ee
Using, for $m =\ell-1$,  
\be
 \left|
 \begin{pmatrix}
 \ell&3&\ell-1\\
 -(\ell-1)&0&\ell-1
 \end{pmatrix}
 \right|^2
 =\frac{6(\ell-1)(2\ell-3)}
 {\ell(\ell+1)(2\ell+1)(2\ell+3)},
 \label{appmm3j}
\ee
we obtain from \eq{RMRM}
\be
 |R_{--}^{(\ell)}|^2 = 
 |F_{\ell,+}|^2
 \frac{8(\ell-1)(\ell+2)(2\ell-3)(2\ell-1)(2\ell+1)}
      {15\ell(\ell+1)}.
      \ee
Finally using \eq{S17R}, we obtain 
\be
 S^{--}_\ell =
 \frac{8(\ell-1)(\ell+2)^2(2\ell-3)(2\ell-1)(2\ell+1)}
 {21(2\ell+3)d_N}[N^2-(\ell+1)^2],
 \ee
 which is Eq.~\eqref{Smm}.
 
The computation for $R_{-+}$ is similar.   Again choosing the convenient
extremal value $m=\ell-1$, with the same incoming coefficients $c_p$
as in \eq{appmmcin}, the outgoing upper-branch
$(\ell+1,j'=\ell+2)$  coefficients are
\be
 d^{(-+)}_{+1}=\sqrt{\frac{\ell(2\ell+1)}{(\ell+2)(2\ell+3)}},\qquad
 d^{(-+)}_{0}=\sqrt{\frac{3(2\ell+1)}{(\ell+2)(2\ell+3)}},\qquad
 d^{(-+)}_{-1}=\sqrt{\frac{3}{(\ell+2)(2\ell+3)}}.
 \label{appmpcout}
\ee
Direct computation as before  gives
\be
 \frac{M^{[1]}_{-+}(\ell-1)}{F_{\ell,+}}=-\mathcal B_\ell,
 \qquad
 \frac{M^{[2]}_{-+}(\ell-1)}{F_{\ell,+}}=\mathcal B_\ell,
 \qquad
 \frac{M^{[3]}_{-+}(\ell-1)}{F_{\ell,+}}=\mathcal B_\ell,
 \label{appmpMpieces}
\ee
where
\be
 \mathcal B_\ell
 :=\frac{4}{\ell+2}
 \sqrt{\frac{15(2\ell-1)}
 {\ell(\ell+1)(2\ell+1)(2\ell+3)}}.
 \label{appBell}
\ee
This time we have
\be
 \left|
 \begin{pmatrix}
 \ell+2&3&\ell-1\\
 -(\ell-1)&0&\ell-1
 \end{pmatrix}
 \right|^2
 =\frac{30}{(\ell+1)(\ell+2)(2\ell+3)(2\ell+5)}.
 \label{appmp3j}
\ee
Equation~\eqref{RMRM} therefore gives
\be
 |R_{-+}^{(\ell)}|^2
 =|F_{\ell,+}|^2
 \frac{8(2\ell-1)(2\ell+5)}
 {\ell(\ell+2)(2\ell+1)}
 \label{appmpR}
\ee
and as a result
\be
 S^{-+}_\ell =
 \frac{40(\ell+1)(2\ell-1)(2\ell+5)}
 {7(2\ell+1)(2\ell+3)d_N}[N^2-(\ell+1)^2],
 \ee
 which is Eq.~\eqref{Smp}.

 For $R_{++}$, we choose $m=\ell+1$. In this case, 
the incoming mode is simple and contains a single term,
\be
 (Y^{\ell,\ell+1,\ell+1})_a=T_{\ell\ell}(e_{+1})_a.
 \label{appppin}
\ee
For the outgoing mode
$(\ell+1,j'=\ell+2,m=\ell+1)$ only two coefficients are nonzero:
\be
 (Y^{\ell+1,\ell+2,\ell+1})_a
 =\sqrt{\frac{\ell+1}{\ell+2}}\,T_{\ell+1,\ell}(e_{+1})_a
 +\frac1{\sqrt{\ell+2}}\,T_{\ell+1,\ell+1}(e_0)_a.
 \label{appppout}
\ee
The finite contraction is therefore especially short. We obtain
\be
\frac{M^{[1]}_{++}(\ell+1)}{F_{\ell,+}}=-(2\ell-1)\mathcal C_\ell,
 \qquad
 \frac{M^{[2]}_{++}(\ell+1)}{F_{\ell,+}}=-2\mathcal C_\ell,
 \qquad
 \frac{M^{[3]}_{++}(\ell+1)}{F_{\ell,+}}=-2\mathcal C_\ell,
\ee
where
\be
 \mathcal C_\ell:=\frac{4}{\ell+2}\sqrt{\frac{\ell}{5}}.
 \label{appCell}
\ee
Thus
\be
 \frac{M_{++}(\ell+1)}{F_{\ell,+}}
 =-(2\ell+3)\mathcal C_\ell.
 \label{appppMfull}
\ee
The corresponding $3j$ factor is
\be
 \left|
 \begin{pmatrix}
 \ell+2&3&\ell+1\\
 -(\ell+1)&0&\ell+1
 \end{pmatrix}
 \right|^2
 =\frac{6(\ell+1)(2\ell+1)}
 {(\ell+2)(\ell+3)(2\ell+5)(2\ell+7)}.
 \label{apppp3j}
\ee
Equation~\eqref{RMRM} then gives
\be
 |R_{++}^{(\ell)}|^2
 =|F_{\ell,+}|^2
 \frac{8\ell(\ell+3)(2\ell+3)^2(2\ell+5)(2\ell+7)}
 {15(\ell+1)(\ell+2)(2\ell+1)}.
 \label{appppR}
\ee
Finally,
\be
S^{++}_\ell = 
 \frac{8\ell^2(\ell+3)(2\ell+3)(2\ell+5)(2\ell+7)}
 {21(2\ell+1)d_N}[N^2-(\ell+1)^2],
\ee
which is Eq.~\eqref{Spp}.

For the fourth kinematically allowed pair,
\be
 j=\ell+1,
 \qquad
 j'=\ell,
\ee
the explicit finite contraction gives
contributions
\be
 R^{[1]}_{+-}:R^{[2]}_{+-}:R^{[3]}_{+-}=-2:1:1.
\ee
Therefore
\be
 R_{+-}^{(\ell)}=0.
 \label{apppmRzero}
\ee
We note that \eq{apppmRzero} is not the result of a selection rule.
It is 
kinematically allowed but dynamically absent due to the specific form of the
vertex $\cV$ for the mode (2,3).

\section{The exceptional (1,0) scaling mode}
\label{scalingmode}

In this appendix, we consider the rotational invariant $(L,K)=(1,0)$
fluctuation, which is a uniform
scaling mode of the fuzzy sphere.  It is exceptional in two related
ways.  First, the factorized part of its one-loop curvature saturates
the positivity inequality and cancels mode by mode, leaving the small
negative curvature coefficient $C_{10}\sim-16/N^2$.  Second, because
the scale coordinate is itself a rotational scalar, the bosonic
one-loop determinant also contains a non-vanishing term \emph{linear}
in the scale coordinate.  By  rotational invariance,
this term is absent for the other  $K>0$ modes. 

The main  result of this appendix is the  expansion of the one-loop
corrected bosonic potential
\be
V_{\rm eff}(q)
=V_{\rm eff}(0)
+\frac{\O_0}{2}A_{10}(N)\,q
+\frac{2\MP}{N^2}\Lambda^{\rm eff}_{10}(N)\,q^2
+\Order(q^3),
\label{master}
\ee
where
\be
A_{10}(N)>0,
\qquad
C_{10}(N)<0,
\qquad
\Lambda^{\rm eff}_{10}
=4-\frac{8a_0}{N}+\Order(N^{-2})>0.
\label{maincoeffs}
\ee
To show this, we start with the
normalized (1,0) vector harmonic
\be
 B^{100}_a
 =
 \frac{J_a}{\sqrt{D_N}},
 \qquad
 D_N:=
 \sum_a\Tr(J_aJ_a)
 =
 \frac{N(N^2-1)}{4}.
 \label{B100}
\ee
The displayed configuration along this mode is therefore
\be
X_a=J_a+qB^{100}_a
=\left(1+\frac{q}{\sqrt{D_N}}\right)J_a
:= \r J_a,
\label{rdef}
\ee
with
\be
\r :=1+\frac{q}{\sqrt{D_N}}
\label{r-q}
\ee
being a scale factor for the fuzzy sphere. In fact under the perturbation,
the radius of the fuzzy sphere becomes
\be
R_q = \r R, \qquad R = N l_P. 
\ee

For the factorized part of the Hessian, we have
\be
\mathsf  a_a:={\rm ad}_{B^{100}_a}
 =
 \frac{1}{\sqrt{D_N}}L_a 
 \label{a100}
\ee
and hence 
\be
\mathsf  b=\frac{1}{\sqrt{D_N}}\,\mathsf d_0 ,
 \label{b-propto-d0}
\ee
or equivalently,
\be
\mathsf  d(q)
 =
\mathsf  d_0+q \mathsf b
 = \r\mathsf  d_0 .
 \label{d-scale}
\ee
Now for
$ \mathsf d_0|r\rangle=\omega_r|u_r\rangle $,
we have
\be
 b_{nr}
 =
 \frac{\omega_r}{\sqrt{D_N}}\,
 \delta_{nr}.
 \label{bnr-scale}
\ee
Thus $b_{nr}=b_{rn}^*$ whenever the matrix element is nonzero,
and $b|r\rangle$ lies entirely in ${\rm im}\,d_0$.
Both inequalities entering the factorization bound are therefore
saturated separately for every internal mode. One can also see this from a
direct mode-by-mode cancellation of the seagull and bubble contributions. In
fact
\be
 \cV_{\rm fact}
 =
 \mathsf d_0^\dagger\mathsf  b+\mathsf b^\dagger \mathsf d_0
 =
 \frac{2}{\sqrt{D_N}}\,\mathsf d_0^\dagger\mathsf  d_0 , \qquad 
 \cW_{\rm fact}
 =
\mathsf  b^\dagger\mathsf  b
 =
 \frac{1}{D_N}\,\mathsf d_0^\dagger \mathsf d_0 .
\ee
Therefore  acting on the mode $|r\rangle$, we have
\be
 (\cV_{\rm fact})_{rr}
 =
 \frac{2\omega_r^2}{\sqrt{D_N}},
 \qquad
 (\cW_{\rm fact})_{rr}
 =
 \frac{\omega_r^2}{D_N}.
 \label{VW-scale}
\ee
The factorized seagull contribution and bubble contribution of this mode is
thus given by
\be
 C^{{\rm fact},r}_{10,\rm sg}
 =
 \frac12
 \frac{(W_{\rm fact})_{rr}}{\omega_r}
 =
 \frac{\omega_r}{2D_N}, \qquad 
 C^{{\rm fact},r}_{10,\rm bub}
 =
 -\frac14
 \frac{|(V_{\rm fact})_{rr}|^2}
 {\omega_r^2(2\omega_r)}
 =
 -\frac{\omega_r}{2D_N}.
\ee
This leads to exact cancellation for every internal mode:
\be \label{Cfact}
C^{{\rm fact}, r}_{10}= C^{{\rm fact},r}_{10,\rm sg}
 + C^{{\rm fact},r}_{10,\rm bub}
 =0 .
\ee
Summing over the internal modes gives
\be
C^{\rm fact}_{10}=0.
\ee
The $(1,0)$ mode therefore  saturates
the factorization inequality as a result of a very special
identity $\mathsf b=\mathsf d_0/\sqrt{D_N}$ which follows from the
uniform-rescaling of the fuzzy sphere.

The complete Hessian contains the lower-derivative remainder.
Under the rescaling \eq{d-scale}, the
eigenvalue $k_n$ of the rescaled fuzzy sphere reads
\be
 k_n(q)
 =
 \r^2 (k_n+2)-2 .
 \label{knexact}
\ee
This determines the bosonic one-loop zero-point energy 
\be
V_B^{(1)}(q)
=\frac{\O_0}{2}\sum_n'\sqrt{k_n(q)}.
\label{V1}
\ee
Expanding Eq.~\eqref{knexact} about $q=0$ gives
\be
\sqrt{k_n(q)}
=\sqrt{k_n}
+\frac{k_n+2}{\sqrt{D_N}\sqrt{k_n}}q
-\frac{k_n+2}{D_N k_n^{3/2}}q^2
+\Order(q^3)
\label{sqrtexp}
\ee
and hence
\be
V_B^{(1)}(q)
=V_B^{(1)}(0)
+\frac{\O_0}{2}A_{10}(N)q
+\frac{\O_0}{2}C_{10}(N)q^2
+\Order(q^3),
\label{V1exp}
\ee
where
\be
A_{10}(N)
:=\frac{1}{\sqrt{D_N}}\sum_n'\frac{k_n+2}{\sqrt{k_n}}>0,\qquad
C_{10}(N)
:=-\frac{1}{D_N}\sum_n'\frac{k_n+2}{k_n^{3/2}}<0.
\label{CA10def}
\ee

To determine these coefficients in the leading order of large $N$, we recall
that there are two branches of $k_n$ entering the sum in \eq{CA10def}: 
\begin{align}
k^-_\ell&=\ell(\ell+3),
&d^-_\ell&=2\ell-1,
&\ell&=2,\ldots,N-1,
\label{minusbranch}\\
k^+_\ell&=(\ell+1)(\ell-2),
&d^+_\ell&=2\ell+3,
&\ell&=3,\ldots,N-1.
\label{plusbranch}
\end{align}
The lower limits are chosen as follows.  The $\ell=1$, $j=0$ mode is
the external $(1,0)$ collective coordinate itself and is therefore
omitted, while the classically non-positive low modes on the other
branch are likewise excluded from the internal determinant.
Therefore we have
\be
A_{10}(N)=\frac{1}{\sqrt{D_N}}\Bigg[
\sum_{\ell=2}^{N-1}(2\ell-1)
\frac{\ell(\ell+3)+2}{\sqrt{\ell(\ell+3)}}+\sum_{\ell=3}^{N-1}(2\ell+3)
\frac{(\ell+1)(\ell-2)+2}{\sqrt{(\ell+1)(\ell-2)}}
\Bigg].
\label{A10sum}
\ee
and 
\be
C_{10}(N)=-\frac{1}{D_N}\Bigg[
\sum_{\ell=2}^{N-1}(2\ell-1)
\frac{\ell(\ell+3)+2}{[\ell(\ell+3)]^{3/2}}+\sum_{\ell=3}^{N-1}(2\ell+3)
\frac{(\ell+1)(\ell-2)+2}{[(\ell+1)(\ell-2)]^{3/2}}
\Bigg].
\label{C10sum}
\ee
The ultraviolet modes of the sums give
\be
  A_{10}(N) = 
\frac{8}{3}N^{3/2} + \Order(N^{-1/2}),
\qquad
C_{10}(N)
 =
 -\frac{16}{N^2}
 +\Order(N^{-3}).
 \label{C10-asymptotic}
\ee
As a result, the one-loop corrected  curvature 
\be
 \Lambda^{\rm eff}_{10}
 =
 4-\frac{8a_0}{N}
 +\Order(N^{-2})
 \label{Lambda10-asymptotic}
\ee
is positive for sufficiently large $N$.
Thus the one-loop-corrected radial curvature remains positive for
sufficiently large N. The physics of the nonvanishing linear $q$-term
will be treated separately.


\section{Numerical support and checks}\label{app:numerics}

The numerical calculations in this appendix are intended as
independent checks of the analytic argument in section 3.4.
Three tests are presented: 1. exhaustive scans of the
complete physical spectrum at finite rank, 2. fixed $x=L/N$ tests of the
large-$N$ expansion of the one-loop coefficient $C_{LK}$,
3. fixed $c=L/\sqrt N$ tests of the crossover expansion
of $C_{LK}$.
Throughout this appendix the
numerical normalization is $a_0=\pi/3$, so that
\be
 \Lambda^{\mathrm{eff}}_{LK}(N)
 =\Lambda^{\mathrm{cl}}_{LK}
 +\frac{\pi N}{6}C_{LK}(N),
 \label{C_num_curvature}
\ee
with
\be
 \Lambda^{\mathrm{cl}}_{L,L-1}=L(L+3),
 \qquad
 \Lambda^{\mathrm{cl}}_{L,L+1}=(L+1)(L-2).
 \label{C_num_classical}
\ee
The longitudinal branch $K=L$ is excluded.  All coefficients quoted
below are exact finite-$N$ one-loop mode-sum coefficients in the
curvature normalization of Sec.~3.1.2.  The detailed numerical
construction and computational workflow are
collected in the final subsection ~\ref{app:num-method}.

\subsection{Exhaustive finite-rank spectra}\label{app:complete-scan}

We have evaluated the complete physical spectrum at $N=30,50,$ and $100$.
We find that at each rank every multiplet on both physical branches
has positive one-loop corrected curvature.
The branch minima are shown in Table~\ref{tab:finiteN-min}.
We find that the global minimum is always given by the $(L,K)=(1,0)$ mode.
Thus the fuzzy sphere has a positive one-loop-corrected Hessian for all of its
physical fluctuation modes.

In  Table~\ref{tab:finiteN-IR} we show the behaviour for 
a number of representative infrared modes.
We note that positivity of the total curvature $\Leff_{LK}$ does not
require a positive $C_{LK}$.  For example, $C_{1,0}<0$ in all three $N$,
while the corrected curvature remains positive.

In  Table~\ref{tab:finiteN-minC}, we list the modes in each branch
with the most negative $C_{LK}$. Note that the corrections are always small
compared to the  order $L^2$ classical curvature.
Note also that negative coefficients also occur in the ultraviolet.

\begin{table}[htbp]
\centering
\caption{Exhaustive finite-rank scans of the static bosonic
  one-loop spectrum.  The number of physical multiplets includes
  both branches $K=L\pm1$.}
\label{tab:finiteN-min}
\begin{tabular}{c c c c c}
\toprule
$N$ & branch & no. of modes scanned & mode of minimum &
$\min\Lambda^{\mathrm{eff}}_{LK}$\\
\midrule
30  & $K=L-1$ & 29 & $(1,0)$ & 3.723513\\
30  & $K=L+1$ & 29 & $(1,2)$ & 9.879147\\
50  & $K=L-1$ & 49 & $(1,0)$ & 3.833258\\
50  & $K=L+1$ & 49 & $(1,2)$ & 18.544347\\
100 & $K=L-1$ & 99 & $(1,0)$ & 3.916392\\
100 & $K=L+1$ & 99 & $(1,2)$ & 39.693306\\
\bottomrule
\end{tabular}
\end{table}

\begin{table}[htbp]
\centering
\caption{Representative infrared coefficients and corrected curvatures in the
 numerical scans.}
\label{tab:finiteN-IR}
\begin{tabular}{c c r r}
\toprule
$N$ & mode $(L,K)$ & $C_{L K}$ & $\Lambda^{\mathrm{eff}}_{LK}$\\
\midrule
30  & $(1,0)$ & -0.017602 & 3.723513\\
30  & $(1,2)$ &  0.756250 & 9.879147\\
30  & $(2,3)$ &  0.844829 & 13.270550\\
50  & $(1,0)$ & -0.006369 & 3.833258\\
50  & $(1,2)$ &  0.784736 & 18.544347\\
50  & $(2,3)$ &  0.937101 & 24.533235\\
100 & $(1,0)$ & -0.001597 & 3.916392\\
100 & $(1,2)$ &  0.796283 & 39.693306\\
100 & $(2,3)$ &  0.977024 & 51.156847\\
\bottomrule
\end{tabular}
\end{table}

\begin{table}[htbp]
\centering
\caption{Most negative  one-loop coefficient on each branch.  Negative
  $C_{LK}$ does not imply instability.}
\label{tab:finiteN-minC}
\begin{tabular}{c c c r r r}
\toprule
$N$ & branch & $L$ & $C_{\min}$ & $\Lambda^{\mathrm{eff}}$ &
$\delta\Lambda/\Lambda^{\mathrm{cl}}$\\
\midrule
30  & $K=L-1$ & 28 & -1.214478 & 848.923031  & -0.02198\\
30  & $K=L+1$ & 27 & -1.812598 & 671.527778  & -0.04067\\
50  & $K=L-1$ & 47 & -1.557085 & 2309.235611 & -0.01735\\
50  & $K=L+1$ & 47 & -1.940393 & 2109.200637 & -0.02352\\
100 & $K=L-1$ & 96 & -1.902378 & 9404.391704 & -0.01048\\
100 & $K=L+1$ & 96 & -2.110551 & 9007.491813 & -0.01212\\
\bottomrule
\end{tabular}
\end{table}

\paragraph{Remark.}
These exhaustive scans provide a direct finite-rank verification of the
positivity of $\Leff_{LK}$ we established in section 3.4.
They do not, by themselves, establish positivity at every larger rank.
That persistence is the subject of the analytic argument in section 3.4.

\subsection{Fixed $x=L/N$ curve}\label{app:fixed-x}

To test the large-$N$ expansion at fixed external momentum fraction $x$,
we consider
\be
 N=30,50,100,200,
 \qquad
 x=0.05,0.10,0.20,0.30,0.50,0.70,0.90,
 \qquad
 L=\operatorname{nint}(xN).
\ee
Here we use the function {\rm nint} to convert $xN$ to the 
nearest integer.
The numerical values of $C_{LK}$ are given in
Tables~\ref{tab:fixedx-Lminus1} and \ref{tab:fixedx-Lplus1} for the minus
and plus branch respectively.
All $56$ corrected curvatures in this scan are positive.

\begin{table}[htbp]
\centering
\caption{Fixed-$x$ values of $C_{L,L-1}(N)$.}
\label{tab:fixedx-Lminus1}
\begin{tabular}{c r r r r}
\toprule
$x$ & $N=30$ & $N=50$ & $N=100$ & $N=200$\\
\midrule
0.05 &  0.792724 &  0.981071 &  1.113557 &  1.187821\\
0.10 &  0.966433 &  1.075503 &  1.107260 &  1.067845\\
0.20 &  1.004938 &  0.921689 &  0.804474 &  0.754667\\
0.30 &  0.694261 &  0.619942 &  0.493878 &  0.420957\\
0.50 &  0.111809 & -0.022343 & -0.193272 & -0.288852\\
0.70 & -0.568872 & -0.764319 & -0.953749 & -1.073814\\
0.90 & -1.213535 & -1.496625 & -1.760177 & -1.919055\\
\bottomrule
\end{tabular}
\end{table}

\begin{table}[htbp]
\centering
\caption{Fixed $x$ values of $C_{L,L+1}(N)$.}
\label{tab:fixedx-Lplus1}
\begin{tabular}{c r r r r}
\toprule
$x$ & $N=30$ & $N=50$ & $N=100$ & $N=200$\\
\midrule
0.05 &  0.844829 &  0.966707 &  1.077053 &  1.134417\\
0.10 &  0.797768 &  0.866571 &  0.920265 &  0.928060\\
0.20 &  0.402208 &  0.411446 &  0.484841 &  0.585385\\
0.30 & -0.193333 &  0.032304 &  0.177670 &  0.256999\\
0.50 & -0.684626 & -0.542197 & -0.475237 & -0.435790\\
0.70 & -1.243522 & -1.220616 & -1.203897 & -1.205327\\
0.90 & -1.812598 & -1.903240 & -1.984318 & -2.037256\\
\bottomrule
\end{tabular}
\end{table}

The main purpose of this scan is not to establish the positivity of
these sampled curvatures, but to test the regular fixed-$x$ expansion
\be
 C_{L,L\pm1}(N)
 =F^{\pm}_0(x)+\frac1N F^{\pm}_1(x)+\frac1{N^2}F^{\pm}_2(x)+\cdots,
 \qquad x=\frac{L}{N}.
 \label{C_fixedx_expansion}
\ee
As a simple convergence diagnostic, we consider
the root-mean-square (RMS) distance
from the $N=200$ curve over $x\ge0.1$.  This is given in
Table~\ref{tab:fixedx-rms}.
The fact that
the RMS is found to decrease monotonically supports the existence of a
finite fixed $x$ limit.

\begin{table}[htbp]
\centering
\caption{RMS deviation from the $N=200$ fixed $x$ curve for $x\ge0.1$.}
\label{tab:fixedx-rms}
\begin{tabular}{c c r}
\toprule
branch & $N$ & RMS deviation\\
\midrule
$K=L-1$ & 30  & 0.420490\\
          & 50  & 0.262301\\
          & 100 & 0.098442\\
\addlinespace
$K=L+1$ & 30  & 0.247379\\
          & 50  & 0.137864\\
          & 100 & 0.058912\\
\bottomrule
\end{tabular}
\end{table}

\paragraph{Remark.}
The fixed $x$ data support the regularity assumption used in
the analytic reasoning: $C_{LK}$ remains bounded and the finite-$N$
curves converge toward a smooth limiting profile.

\subsection{Fixed $c$ curve for the crossover $L=c\sqrt N$}\label{app:sqrtn}

We next test directly the crossover region by considering
\be
 L=\operatorname{nint}(c\sqrt N),
 \qquad
 c=0.5,1,1.5,2,3,
 \qquad
 N=20,30,50,60,100,200.
\ee
The obtained $C$-coefficients are shown in Tables~\ref{tab:sqrtn-Lplus1}
and \ref{tab:sqrtn-Lminus1}.  There are nine negative values of $C$
among the $60$ data points. They all occur at the smaller sampled $N$,
where the finite-$N$ corrections proportional to $L/N=c/\sqrt{N}$
are still appreciable.

\begin{table}[htbp]
\centering
\caption{Values of $C_{L,L+1}(N)$ for $L=\operatorname{nint}(c\sqrt N)$.}
\label{tab:sqrtn-Lplus1}
\begin{tabular}{c r r r r r r}
\toprule
$c$ & $N=20$ & $N=30$ & $N=50$ & $N=60$ & $N=100$ & $N=200$\\
\midrule
0.5 &  0.673635 &  0.797768 &  0.944276 &  0.994795 &  1.077053 &  1.157217\\
1.0 &  0.463521 &  0.515064 &  0.692597 &  0.750410 &  0.920265 &  1.062570\\
1.5 & -0.516965 &  0.020546 &  0.287806 &  0.410245 &  0.678991 &  0.902946\\
2.0 & -0.706771 & -0.344713 &  0.076128 &  0.209769 &  0.484841 &  0.764622\\
3.0 & -1.090892 & -0.784436 & -0.336103 & -0.182180 &  0.177670 &  0.549520\\
\bottomrule
\end{tabular}
\end{table}

\begin{table}[htbp]
\centering
\caption{Crossover values of $C_{L,L-1}(N)$ for $L=\operatorname{nint}(c\sqrt N)$.}
\label{tab:sqrtn-Lminus1}
\begin{tabular}{c r r r r r r}
\toprule
$c$ & $N=20$ & $N=30$ & $N=50$ & $N=60$ & $N=100$ & $N=200$\\
\midrule
0.5 &  0.784276 &  0.966433 &  1.054448 &  1.060814 &  1.113557 &  1.173940\\
1.0 &  0.996127 &  1.017506 &  1.051394 &  1.064007 &  1.107260 &  1.157390\\
1.5 &  0.646503 &  0.831258 &  0.842005 &  0.876601 &  0.963064 &  1.048591\\
2.0 &  0.406075 &  0.535448 &  0.668056 &  0.717015 &  0.804474 &  0.931393\\
3.0 & -0.199464 & -0.008480 &  0.221142 &  0.302840 &  0.493878 &  0.718969\\
\bottomrule
\end{tabular}
\end{table}

The analytic fixed-$x$ expansion implies, under the assumption that the
coefficient functions $F_n^\pm(x)$ are smooth at the endpoint, 
\be
 C_{L,L\pm1}(N)
 =F_{\pm}+a_{\pm}\frac{c}{\sqrt N}+O(N^{-1})
 =F_{\pm}+a_{\pm}\frac{L}{N}+O(N^{-1}),
 \label{C_sqrtn_expansion}
\ee
where the leading coefficient $F_{\pm}$ is independent of every fixed
finite $c$.
To test
\eq{C_sqrtn_expansion}, we fit the $N\ge100$ data to
\be
 C_i=F+a z_i,\qquad z_i=\frac{L_i}{N_i},
 \label{C_two_parameter_fit}
\ee
using the actual integer $L_i=\operatorname{nint}(c_i\sqrt{N_i})$
employed in the numerical calculation. There are ten data points
(five $c$ values at each of $N=100$ and $200$)
and two fit parameters, $F$ and $a$.  An unweighted least-squares fit gives
\begin{align}
 K=L+1:&\qquad C\simeq1.2854-3.7339\frac{L}{N},
 \qquad \mathrm{RMS}=0.0311,\qquad R^2=0.9888,
 \\
 K=L-1:&\qquad C\simeq1.3091-2.6309\frac{L}{N},
 \qquad \mathrm{RMS}=0.0400,\qquad R^2=0.9634.
 \label{C_two_parameter_results}
\end{align}
Here RMS denotes the root-mean-square residual
$ r_i :=C_i-(F+aL_i/N_i)$ and
$R^2$ is the coefficient of determination, which
measures how much of the variation of the data is explained by the
fitted model. The quality of these two-parameter fits provides
strong numerical support  for the predicted linear $L/N$
dependence of the leading crossover behaviour.

\paragraph{Remark.}
The crossover data support the feature that is most important
for the analytic argument in
section 3.4: for fixed finite $c$, the leading large-$N$ coefficient
is compatible with a positive $c$-independent intercept,
with a linear $c$-dependence entering through finite-$N$ corrections.

\subsection{Band representation and numerical construction}
\label{app:num-method}

This subsection records how the numbers in the preceding tables are
obtained.  It is placed last so that the numerical results and their
role in the stability argument can be read independently of
implementation details.  The performed 
calculation is a finite-$N$
numerical evaluation of the one-loop mode sum.

For numerical computation on the fuzzy sphere, it is convenient to use
a band representation of matrices
whose nonzero entries lie only on a small number of diagonals
parallel to the main diagonal. 
Storing and manipulating only these nonzero bands substantially
reduces both memory requirements and computational cost, without introducing
any approximation.
An example is the scalar fuzzy harmonics defined by
\be
 \sum_{a=1}^{3}[J_a,[J_a,T_{\ell m}]]
 =\ell(\ell+1)T_{\ell m},
 \qquad
 \operatorname{Tr}(T_{\ell m}^{\dagger}T_{\ell' m'})
 =\delta_{\ell\ell'}\delta_{mm'}.
 \label{C_scalar_harmonics}
\ee
In the $J_3$-diagonal basis, $[J_3, T_{\ell m}] = m T_{\ell m}$ implies that 
$T_{\ell m}$ has support on a single matrix diagonal.

The numerical calculation  begins with the construction of the
scalar fuzzy harmonics $T_{\ell m}$.
For fixed $m$, the adjoint Laplace operator $[J_a, [J_a, \cdot]]$
is tridiagonal.  Diagonalizing
this tridiagonal operator gives the corresponding
band arrays for $T_{\ell m}$.
Once the scalar harmonics are known, the external vector harmonics
$ B_a^{LKQ}$ and the internal vector harmonics $Y_a^{\ell j m}$
are constructed from them by
Clebsch–Gordan  coupling to the spin-one vector index. The
displaced-Hessian vertices $\cV$ and $\cW$ are then evaluated
directly in the band representation, and the required mode-space matrix elements
$\cV_{nr}$ and $\cW_{nn}$ are used to compute the mode sum \eq{CLK}.

 \section*{Acknowledgments}

 We thank  Harold Steinacker for discussions and comments.
We acknowledge the support of this work by NCTS, the National Science and
Technology Council of Taiwan for the grant 113-2112-M-007-039-MY3, and
the National Tsing Hua University 2025 Talent Development Fund for a
TSAI WANG, YUAN-YANG Distinguished Talent Chair Professorship.
Numerical computations and symbolic manipulations, including the
construction of matrix operators, evaluation of mode sums, parameter
scans, and data fitting, were carried out with assistance from OpenAI
ChatGPT and Codex. All derivations, computational procedures, and
results were independently checked by the author, who takes full
responsibility for the content of this work.

\bibliographystyle{utphys}

\bibliography{references}  
\end{document}